\documentclass[aps,prd,groupedaddress,nofootinbib,superscriptaddress,
preprint,11pt,a4paper]{revtex4-1}
\usepackage[colorlinks=true,citecolor=magenta,linkcolor=blue,
breaklinks=true]{hyperref}
\usepackage{amsmath,amssymb}
\usepackage{graphicx,graphics}
\usepackage[caption=false,labelformat=simple]{subfig}
\usepackage{slashed}
\usepackage{orcidlink}
\newcommand{\SplitFootnoteMark}[1]{%
	\ensuremath{^{\rm\ref{#1}}}%
}

\newcommand{\SplitFootnoteText}[2]{%
	\refstepcounter{footnote}%
	\footnotetext[\value{footnote}]{\label{#1} #2}%
}
\begin{document}
%%%%%%%%%%%%%%%%%%  Title %%%%%%%%%%%%%%%%%%%%%%%%%%%%%%%%%%%%%%%%%%%%%%
\title{Search for Lepton Flavor Violating Signals at the Future
Electron-Proton Colliders}

%%%%%%%%%%%%%%%%%%  Authors  %%%%%%%%%%%%%%%%%%%%%%%%%%%%%%%%%%%%%%%%%%%
\author{Anjan Kumar Barik\,\orcidlink{0000-0001-6649-9033}}
\email{anjanbarik@hri.res.in}
\affiliation{Regional Centre for Accelerator-based Particle Physics,
Harish-Chandra Research Institute, A CI of Homi Bhabha National
Institute, Chhatnag Road, Jhunsi, Prayagraj 211\,019, India.}

\author{Atri Dey\,\orcidlink{0000-0002-1645-7641}}
\email{atridey1993@gmail.com}
\affiliation{School of Theoretical Physics, Dublin Institute for
Advanced Studies, 10 Burlington Road, Dublin 4, Ireland}

\author{Tousik Samui\,\orcidlink{0000-0002-1485-6155}\,}
\email{tousiks@imsc.res.in}
\email{tousiksamui@gmail.com}
\affiliation{Department of Physical Sciences, Indian Institute of
Science Education and Research Kolkata,\\ Mohanpur 741\,246, India.}
\affiliation{The Institute of Mathematical Sciences,\\ IV Cross Road,
CIT Campus, Taramani, Chennai 600\,113, India.}
%%%%%%%%%%%%%%%%%%%%%%%%%%%%%%%%%%%%%%%%%%%%%%%%%%%%%%%%%%%%%%%%%%%%%%%%

\preprint{DIAS-STP-23-10}
\preprint{HRI-RECAPP-2023-02}

\begin{abstract}
The search for lepton flavor violation (LFV) is a powerful probe to look
for new physics beyond the Standard Model. We explored the possibility
of searches for LFV $Z$ boson couplings to electron and muon pairs at
the upcoming electron-proton colliders, namely the Large Hadron Electron
Collider (LHeC) and the Future Circular lepton-hadron Collider (FCC-eh).
We employed the study via a single muon plus an associated jet channel
to search for the LFV signal. We used a multivariate technique to obtain
an improved signal-background analysis. By using the condition on
nonobservation of any significant deviation of the signal over the
expected background, we provide an upper limit on the LFV $Z$ boson
coupling and corresponding branching ratio (BR). We find that an upper
limit of up to $1.13\times 10^{-7}$ and $4.64 \times 10^{-8}$ can be set
on BR($Z\to e \mu$) at 95\% confidence level (C.L.) with one year run of
LHeC and FCC-eh, respectively, if the LFV coupling is governed by vector
or axial-vector coupling. For tensor or axial-tensor coupling, these
limits can be improved up to $2.34\times 10^{-8}$ and
$5.02\times 10^{-9}$ for LHeC and FCC-eh machines, respectively, at 95\%
C.L. The projected numbers improve significantly over the existing limit
of $2.62\times 10^{-7}$ set by ATLAS.
\end{abstract}

\maketitle

%%%%%%%%%%%%%%%%%%%%%%%%%%%%%%%%%%%%%%%%%%%%%%%%%%%%%%%%%%%%%%%%%%%%%%%%
\section{Introduction}
The successful framework of the Standard Model (SM) of particle physics
respects lepton flavor conservation, although the observations of
neutrino masses and mixing\,\cite{SNO:2002tuh,Super-Kamiokande:1998kpq}
provides non-zero lepton flavor violation (LFV)\footnote{The phrase
`lepton flavor violating' is also abbreviated as LFV in this article.
The exact meaning of the abbreviation should be clear from the context.}
via the loop. The amount of such violations is extremely small to be
detected in an experiment. To date, no experimental measurement shows a
piece of direct evidence in support of charged
LFV\,\cite{Calibbi:2017uvl,ParticleDataGroup:2022pth}. However, the
violation of the lepton flavor always remains a topic of interest in the
particle physics community. Primarily because an experimental
observation of establishing lepton flavor violation opens up a plethora
of avenues of new physics beyond the Standard Model (BSM). For example,
various neutrino mass models, such as Type-II, Type-III, and inverse
seesaw models, which generate neutrino mass at the tree level, exhibit
LFV scenarios. On the other hand, some models having neutrino mass
generated radiatively, {\it e.g.} Zee-Babu model, Scotogenic models,
etc., also show signatures for
LFV\,\cite{Konetschny:1977bn,Foot:1988aq,Sun:2013kga,Mohapatra:1980yp,Hisano:1995nq,Herrero-Garcia:2014usa,Babu:1988ki,Zee:1980ai,Ma:2006km,Rocha-Moran:2016enp,Hundi:2022iva}.

A series of experiments, both dedicated and general-purpose, have been
performed at various levels to search for LFV in the charged leptons.
None of which shows any significant evidence for it. This, in turn, sets
upper limits to the LFV branching ratios (BRs) and LFV couplings of the
decays of various particles. The MEG experiment sets an upper limit on
BR($\mu\to e \gamma$) at $4.2\times 10^{-13}$\,\cite{MEG:2016leq}, the
SINDRUM experiment sets an upper limit on BR($\mu\to 3e$) at
$1.0\times 10^{-12}$\,\cite{SINDRUM:1987nra}, the BaBar experiment sets
an upper limit on BR$(\tau\to e\gamma)$ at $3.3\times 10^{-8}$ and on
BR$(\tau\to\mu\gamma)$ at $4.4\times 10^{-8}$\,\cite{BaBar:2009hkt}, the
Belle experiment sets a limit on BR$(\tau\to 3e)<2.7\times 10^{-8}$ and
BR$(\tau\to3\mu)<4.4\times 10^{-8}$\,\cite{Hayasaka:2010np} at 90\%
confidence level (C.L.) at all the experiments.

Another series of bounds is given on the LFV decays of heavy neutral
bosons. These types of bounds primarily come from the collider
experiments because of their ability to produce on-shell heavy bosons.
The LFV decays of such particles have been searched for at the Large
Hadron Collider (LHC). The limits on the BRs of the Higgs boson in the
$e\mu$, $e\tau$, and $\mu\tau$ channels have been set to be no more than
$3.5\times 10^{-4}$\,\cite{CMS:2016cvq}, $6.1\times 10^{-3}$, and
$2.5\times 10^{-3}$\,\cite{CMS:2017con}, respectively at 95\% C.L.~by
CMS collaboration at the LHC. On the other hand, the searches for LFV
$Z$ decays (LFVZD) date back to the era of the Large Electron-Positron
(LEP) collider, where the neutral boson could be produced copiously. The
LFV decays of the neutral boson $Z$ have been measured in $e \mu$,
$e\tau$, and $\mu\tau$ channels in LEP by OPAL and DELPHI
collaborations\,\cite{OPAL:1995grn,DELPHI:1996iox}. However, a recent
search of LFVZD in the $e\mu$ channel at the LHC by the ATLAS
collaboration\,\cite{ATLAS:2022uhq} sets the limit on BR($Z\to e\mu$) at
$2.62\times 10^{-7}$ at 95\% C.L., which supersedes the previous bound
by LEP.

Many in-depth studies, both theoretical and experimental, of various LFV
signals have been conducted in the context of lepton and hadron
colliders. In the Hadro-Electron Ring Accelerator (HERA), LFV signals
were looked at by H1 and ZEUS collaborations in the context of
leptoquark
searches\,\cite{ZEUS:1996bbn,H1:1999dil,ZEUS:2002dnh,ZEUS:2005nsy,H1:2007dum,H1:2011rlk}.
The potential of electron-proton colliders like the Future Circular
Collider (FCC-eh)\,\cite{FCC:2018byv} and the Large Hadron electron
Collider
(LHeC)\,\cite{LHeCStudyGroup:2012zhm,Bruening:2013bga,LHeC:2020van},
however, has not been covered fully in any of the literature that has
been published thus far. Among these, the LHeC stands out owing to its
capacity to function concurrently with the HL-LHC, made possible by the
establishment of the energy recovery linac, especially for the electron
beam\,\cite{LHeC:2020van}. Although the major objective of
electron-proton colliders is to give high-precision data that enables
precise determination of the parton distribution functions of a proton,
their special benefits over pp colliders in identifying rare BSM events
have not been adequately recognized. The use of an electron beam can
curb underlying events, pile-ups, or QCD backgrounds in particular. When
examining rare BSM events, which may easily be wiped out by the
overwhelming presence of more than 150 pile-ups at the HL-LHC, the
insignificant pile-up environment of LHeC (FCC-eh) being roughly 0.1 (1)
pile-up collisions per event proves to be quite helpful. One also can
anticipate strong background control for electron-proton colliders by
separating charged-current (CC) and neutral-current (NC) processes, as
well as determining the forward direction. Therefore, the study of LFV
could be explored in more detail in the upcoming electron-proton
colliders.

The projections for such LFV decays have also been studied in the future
electron-positron
colliders\,\cite{Etesami:2021hex,Calibbi:2021pyh,Altmannshofer:2023tsa,Dev:2017ftk}.
Similar studies have also been performed in the future electron-proton
colliders\,\cite{Alan:2001cf,Gonderinger:2010yn,Accardi:2012qut,Cirigliano:2021img,AbdulKhalek:2021gbh,AbdulKhalek:2022hcn,Jueid:2023fgo}.
However, these studies primarily focused on the electron-tau LFV
scenarios. In this work, we investigate the discovery potential of LFV
in future electron-proton colliders, namely the LHeC and the FCC-eh. We
have primarily looked at the possibility of measuring BR($Z\to e\mu$) at
these upcoming colliders. For this, we employed a search strategy via
$\mu + j$ channel, where a single $\mu$ is produced due to the LFV
coupling with the $Z$ boson. This channel can provide good sensitivity
depending on the nature of the lepton-$Z$ coupling boson -- vector,
axial-vector, tensor, or axial-tensor. Our study suggests that at least
a comparable result to the LHC can be obtained in the LHeC, and an
improved result can be expected at the FCC-eh machine.

The outline of this article is as follows. In Section~\ref{sec:model},
we briefly discuss the generic Lagrangian providing LFV coupling to the
$Z$ boson. We describe our analysis strategy in
Section~\ref{sec:method}. We then discuss the methods and multivariate
analysis in Section~\ref{sec:mva}. We present our results in
Section~\ref{sec:result}. After that, we summarize our findings in
Section~\ref{sec:summary}.

%=======================================================================
\section{Prospect of the search for LFV $Z$ boson coupling at
electron-proton collider}\label{sec:model}\vspace{-10pt}
We study the LFVZD in a model-independent way. For this, we consider the
following general-purpose BSM Lagrangian.
\begin{equation}
\mathcal{L}^\text{eff}_Z = \bar\ell_i \gamma^\mu (g_v^{ij}
    + g_{av}^{ij}\gamma_5) \ell_j Z_\mu + \bar\ell_i \sigma^{\mu\nu}
     (g_t^{ij} + g_{at}^{ij}\gamma_5) \ell_j Z_{\mu\nu} + h.c.,
\label{eqn:lag}
\end{equation}
where $\ell_i$ represents the $i^\text{th}$ generation lepton and
$g^{ij}$'s are the coupling constants of $\ell_i$ and $\ell_j$ pair with
the $Z$ boson, considering only the BSM contribution. In the Lagrangian,
the indices $i$ and $j$ may be equal or different; $i\neq j$ corresponds
to LFV. The subscripts of $g^{ij}$ indicate the Lorentz structure,
namely vector ($v$), axial-vector ($av$), tensor ($t$), or axial-tensor
($at$), of the couplings with the $Z$ boson. This Lagrangian should not
be identified as an ultraviolet complete model. In the popular BSM
models, these types of interaction are usually generated when the
particles with non-diagonal flavor coupling to the lepton sector run in
the loop\,\cite{Illana:2000ic,DeRomeri:2016gum}. In the flavor
conserving models, the flavor off-diagonal couplings of the $Z$ boson
with the leptons are absent in the tree-level Lagrangian, thereby
avoiding the ultraviolet divergences via the loops. In such
flavor-conserving models, the effective LFV couplings arise only through
loop-level processes and are typically subject to symmetry constraints
(such as gauge invariance or flavor symmetries). These constraints often
lead to relations among the different $g^{ij}$ couplings.

In terms of these couplings, the BR of the $Z$ boson to $\ell_i \ell_j$
pair becomes
\begin{equation}
\text{BR}(Z\to \ell_i \ell_j) = \frac{M_Z}{12\,\pi\,\Gamma_Z^\text{SM}}
     \Big[2(|C_v^\ell\delta_{ij} + g_v^{ij}|^2 + |C_{av}^\ell\delta_{ij}
    + g_{av}^{ij}|^2) + M_Z^2(|g_t^{ij}|^2 + |g_{at}^{ij}|^2)\Big],
\label{eqn:brZ}
\end{equation}
where $C_v^\ell$ and $C_{av}^\ell$ are the SM vector and axial-vector
couplings of the $Z$ boson with the charged leptons, respectively.
Please note that $g^{ij}$ represents any contribution arising solely
from BSM physics. In Eq.~(\ref{eqn:brZ}), we assumed that the
final-state leptons are massless and set the kinematic phase-space
factor to unity, thereby neglecting any threshold effects in the decay.
Furthermore, we also assumed that the total decay width of the $Z$ boson
remains approximately equal to its SM value, since the BSM contributions
to the total width are negligible. The upper limit on the BR of LFVZD
can be translated to the couplings in the Lagrangian in
Eq.~(\ref{eqn:lag}). In Table~\ref{tab:Zbounds}, we tabulate the current
upper limits on various LFV $Z$ boson BRs, along with the corresponding
bounds on the BSM coupling constants derived from them. The
corresponding bounds on the BSM coupling constants are obtained assuming
that only one coupling is non-zero at a time, while the others are set
to zero.

%=======================================================================
\begin{table}[!h]\renewcommand{\arraystretch}{1.25}
\begin{center}
\begin{tabular}{|c|c|c|c|c|} 
\hline
  & Experiment & BR & $g_v^{ij}$ = $g_{av}^{ij}$ & $g_t^{ij}$ = $g_{at}^{ij}$ \\
\hline
 $Z\to e \mu$  & ATLAS\,\cite{ATLAS:2022uhq} & $2.62\times 10^{-7}$ & $3.68\times 10^{-4}$ & $5.71\times 10^{-6}$~GeV$^{-1}$ \\ 
\hline
 $Z\to e \tau$  & ATLAS\,\cite{ATLAS:2021bdj} & $5.0\times 10^{-6}$ & $1.61\times 10^{-3}$ & $2.49\times 10^{-5}$~GeV$^{-1}$ \\ 
\hline
 $Z\to \mu \tau$ & ATLAS\,\cite{ATLAS:2021bdj} & $6.5\times 10^{-6}$ & $1.83\times 10^{-3}$ & $2.84\times 10^{-5}$~GeV$^{-1}$\\ 
\hline
\end{tabular}
\caption{Current upper limit at 95\% C.L.~on the LFV $Z$ boson BRs. The
upper limit on various couplings derived from the bounds on the LFV $Z$
BRs [Eq.~(\ref{eqn:brZ})]. A single coupling has been considered to be
nonzero while deriving the bounds.}
\label{tab:Zbounds}
\end{center}
\vspace{-5mm}
\end{table}
%=======================================================================
\vspace{-20pt}
\subsection{Signals and backgrounds} \label{sec:method}
\begin{figure}[h]
\begin{center}
	\includegraphics[width=0.40\textwidth]{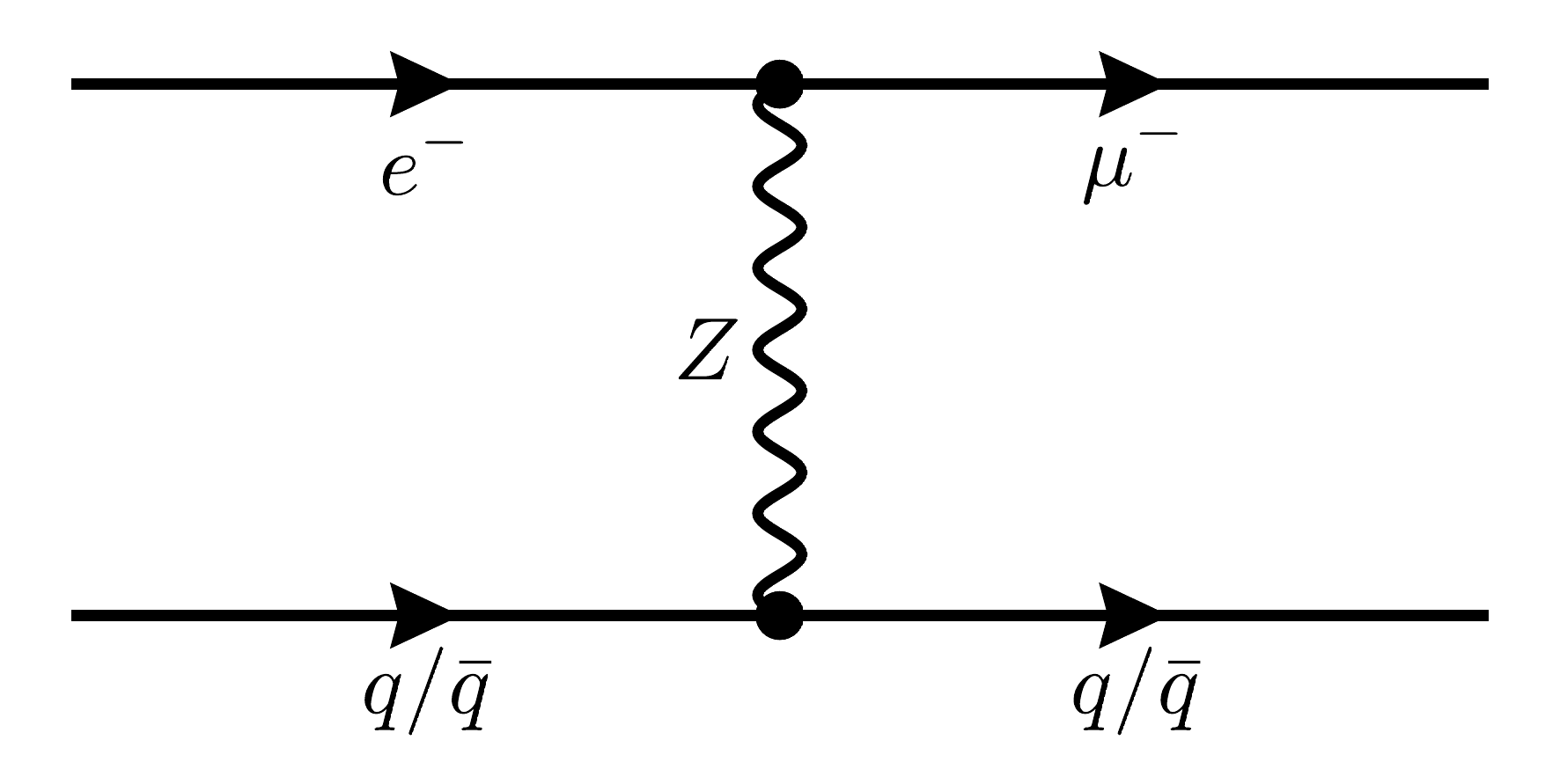}
\end{center}
\vspace{-30pt}
\caption{Feynman diagram for the signal process $e^- p \to \mu^- j$.}
\label{fig:signal-feyndia}
\vspace{-5pt}
\end{figure}

\vspace{-10pt}
In order to study the LFV scenario in an electron-proton collider, we
consider the process
\begin{eqnarray}
e^- p \to \mu^- j.
\end{eqnarray}
The LFV couplings introduced in the Lagrangian in Eq.~(\ref{eqn:lag})
would induce such a process. The Feynman diagram of the process is shown
in Fig.~\ref{fig:signal-feyndia}.
%=======================================================================
%=======================================================================
In any electron-proton collider, the production of a single muon without
any source of missing energy is unlikely in the SM because of the
conservation of the lepton number of each generation. Therefore, this
provides an interesting channel to search for lepton flavor violations.
If no signal for such a violation is observed at the collider, one may
put a constraint on the LFV couplings introduced in Eq.~(\ref{eqn:lag}),
namely $g^{e\mu}_v$, $g^{e\mu}_{av}$, $g^{e\mu}_t$, and $g^{e\mu}_{at}$.
We can then use this bound on the LFV couplings to provide an upper
limit on the BRs of the LFVZD using Eq.~(\ref{eqn:brZ}). In this work,
we will be considering one type of coupling out of the $g^{e\mu}_v$,
$g^{e\mu}_{av}$, $g^{e\mu}_t$, or $g^{e\mu}_{at}$, that is to say, one
of vector, axial-vector, tensor, or axial-tensor operators at a time. We
have deferred the scope of the study containing more than one operator
to a separate work.

We have considered the LHeC and the FCC-eh for the searches for the
above signal. Each machine has been proposed to run in two phases. We
label these phases as `LHeC1', `LHeC2', `FCC-eh1', and `FCC-eh2' for
further reference. We also list the most important specifications of
these machines in Table~\ref{tab:coll-param}.
%\newpage
%=======================================================================
\begin{table}[h]\renewcommand{\arraystretch}{1.25}
\begin{center}
\begin{tabular}{|l|r|r|r|r|}
\hline & \quad LHeC1\,\cite{LHeCStudyGroup:2012zhm} \qquad & \quad LHeC2\,\cite{Bruening:2013bga} \qquad & \quad FCC-eh1\,\cite{LHeC:2020van} \qquad & \quad FCC-eh2\,\cite{FCC:2018byv} \qquad \\
\hline ~Electron energy [GeV]  & 30 & 50 & 60 & 60 \\
\hline ~Proton energy [GeV]   & 7000 & 7000  & 20000 & 50000 \\
\hline ~Luminosity [nb$^{-1}$s$^{-1}$] & 5 & 9 & 8  & 15  \\
\hline ~Integrated Luminosity (1 year run) [fb$^{-1}$] & 50 & 90 & 80 & 150 \\
\hline
\end{tabular}
\caption{Some important specifications of the upcoming electron-proton
colliders. For the integrated luminosity, we have assumed a run time of
$10^7$ seconds per year\SplitFootnoteMark{tablefootnote}.}
\label{tab:coll-param}
\end{center}
\vspace{-16pt}
\end{table}
\SplitFootnoteText{tablefootnote}{This is a conservative estimate for
effective data-taking time at collider experiments after accounting for
downtime and various operational constraints.}
%=======================================================================

\begin{figure}[h!]
\centering
\subfloat[]{\includegraphics[width=0.48\textwidth]{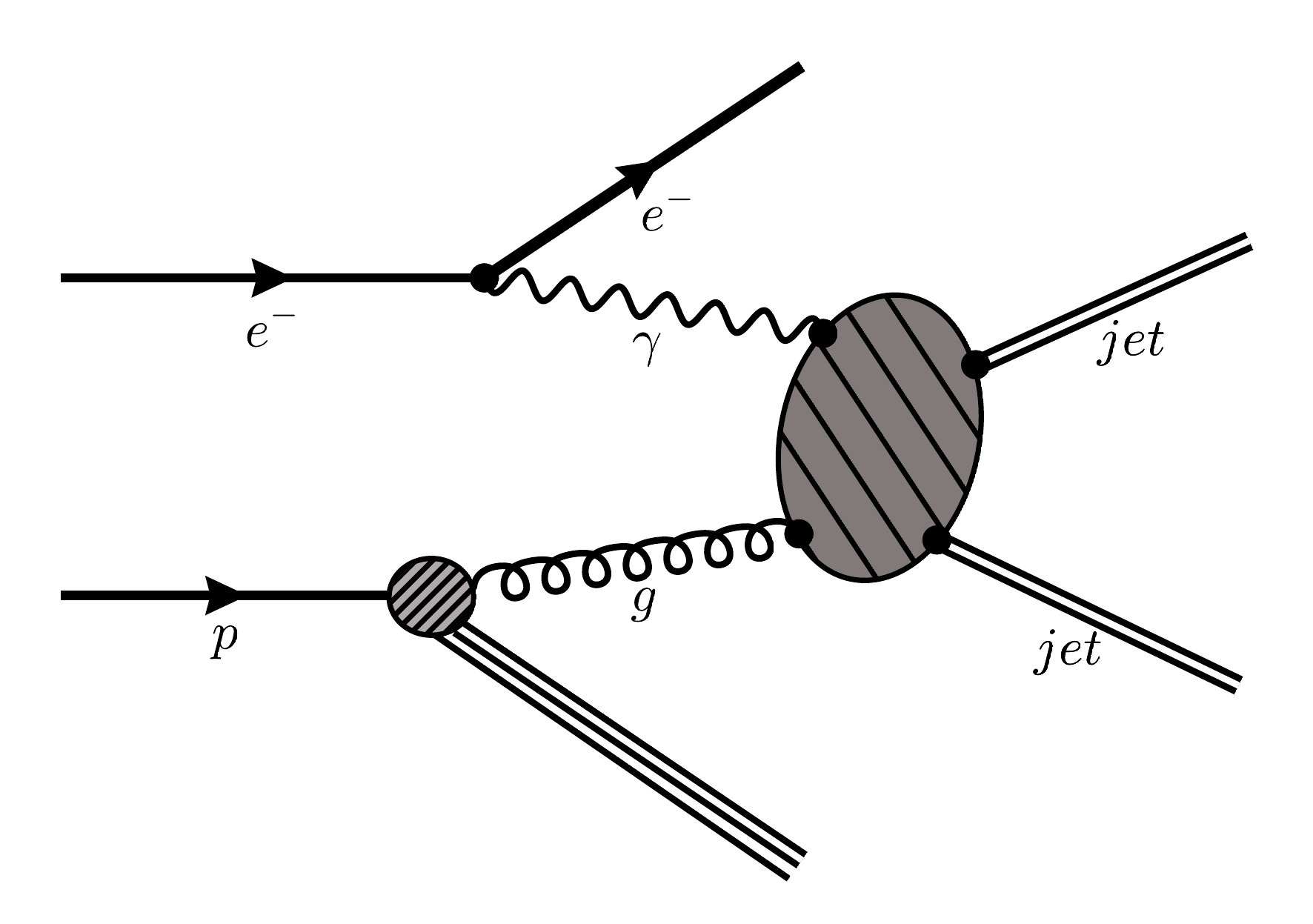}\label{fig:feyndia:a}}
\subfloat[]{\includegraphics[width=0.40\textwidth]{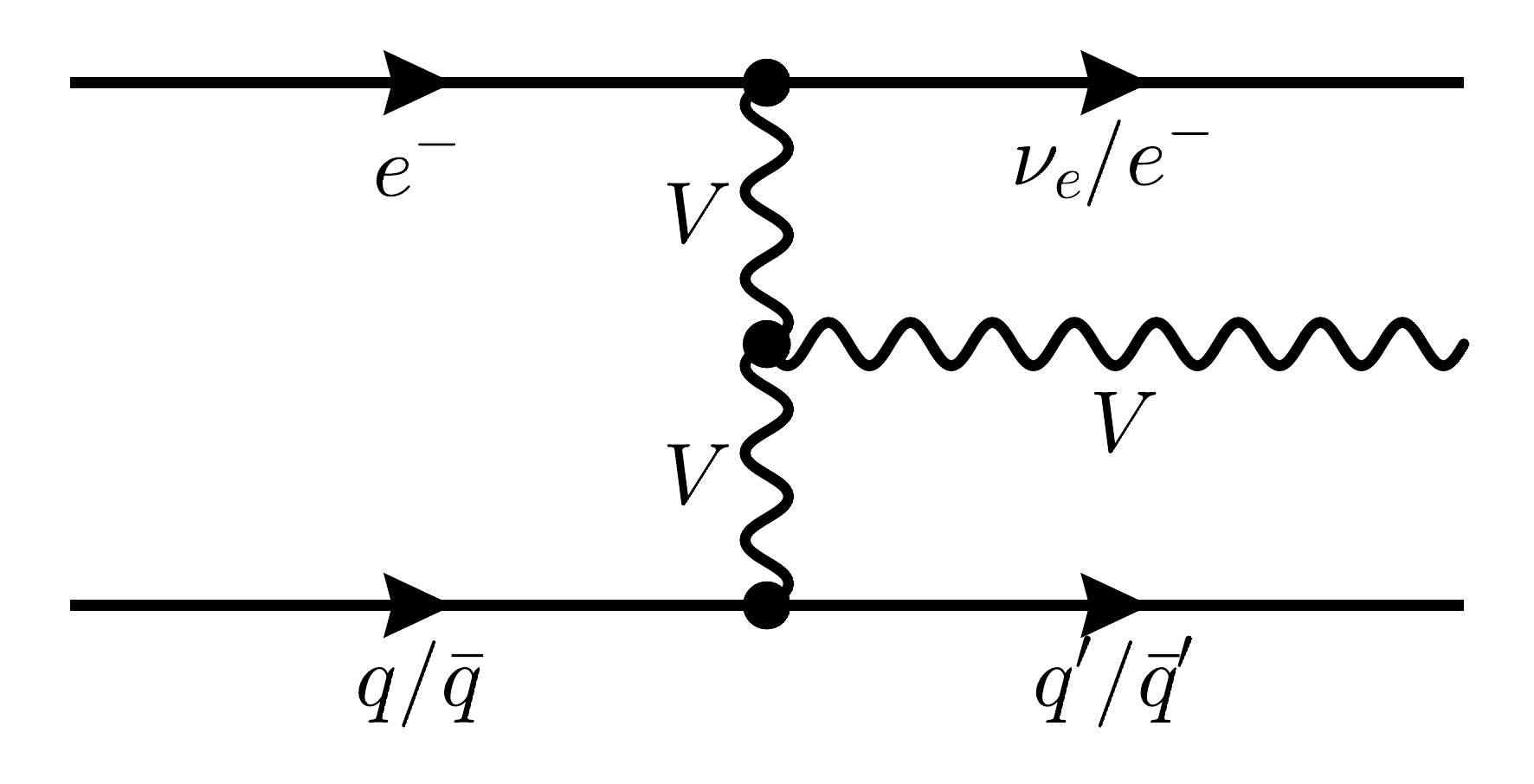}\label{fig:feyndia:b}}
\vspace{-0.0cm}
\caption{Representative Feynman diagrams for (a) the photoproduction
process and (b) other background processes with $V = W^\pm, \gamma, Z$.}
\label{fig:feyndia}
\end{figure}

In an electron-proton collision, the signal process arises due to a
lepton flavor-changing interaction mediated by a $Z$ boson exchanged in
the $t$-channel. This particular channel is not present in the SM.
However, some other SM processes can give rise to the same final state
as the signal. The following SM processes are considered to be potential
backgrounds for our signal.
\begin{center}
\begin{tabular}{r l l l}
\textbf{Photoproduction:} & \quad $e^- p$ & $\longrightarrow e^- \gamma p$ & $\longrightarrow e^- X$ \\
\textbf{mu:} & \quad $e^- p$ & $\longrightarrow \nu_e j W$ & $\longrightarrow \nu_e j \mu \nu_\mu$ \\
\textbf{e-mu:} & \quad $e^- p$ & $\longrightarrow e^- j W$ & $\longrightarrow e^- j \mu \nu_\mu$   \\
\textbf{mu-mu:}    & \quad $e^- p$ & $\longrightarrow \nu_e j \gamma^*/Z$ & $\longrightarrow \nu_e j \mu \mu$ \\
\end{tabular}
\end{center}
Some representative Feynman diagrams for these background processes are
shown in Fig.~\ref{fig:feyndia}.
%=======================================================================
%=======================================================================

\begin{itemize}
\item \textbf{Photoproduction:} The primary background comes from
photoproduction\,\cite{Bauer:1977iq,Chwastowski:2003aw,ZEUS:2005nsy,H1:2011rlk},
wherein direct or resolved photons from the electron beam interact with
the proton beam. The hadronic final state can give rise to muons from
the (semi)leptonic decays of heavy $D$ or $B$ mesons.\\[-30pt]

\item \textbf{mu:} Another important background comes from the
production of a $W$ boson and a jet along with an invisible neutrino.
The $W$ boson can then decay into a muon and a neutrino. This background
contains a single muon and has the same visible final state as that of
the signal. However, the large missing energy from the invisible
neutrinos makes the background reduction easier.\\[-30pt]

\item \textbf{e-mu:} The third main background primarily comes from the
production of an electron, a jet, and a $W$ boson. The subsequent decay
of the $W$ boson to the muon channel gives a similar final state as the
signal if the electron in the final state gets missed. The source of
missing energy makes the background reduction easier.\\[-30pt]

\item \textbf{mu-mu:} The fourth background is similar to the first one,
but involves the production of a $\gamma^*/Z$ boson along with the jet
and MET. The decay of $\gamma^*/Z$ to two muons and one of them being
missed makes its final state very similar to that of the
signal.\\[-30pt]
\end{itemize}
We further note that the same event topology in the backgrounds also
appears when the vector bosons decay to $\tau$ lepton(s) and the $\tau$
then decays leptonically. We have taken care of this aspect in the
background event generation. Furthermore, top quark production and
semileptonic decays of heavy-flavor quarks can also mimic the signal
topology and act as potential backgrounds. However, we have checked that
the contributions from these processes are negligible.
%=======================================================================
\begin{table}[b]
\centering
\begin{tabular}{|c|c|c|c|c|}
\hline
\textbf{} & \textbf{e-mu [fb]} & \textbf{mu-mu [fb]} & \textbf{mu [fb]}& \textbf{Photoproduction [fb]}\\
\hline
LHeC1   &  174 & 12.1  & 57.2 & $5.48 \times 10^4$\\
LHeC2   &  294 & 20.6  & 100  & $7.77 \times 10^4$\\
FCC-eh1 &  758 & 53.5  & 284  & $1.68 \times 10^{5}$ \\
FCC-eh2 & 1340 & 93.6  & 519  & $2.84 \times 10^{5}$\\
\hline
\end{tabular}
\caption{Cross sections for background processes across different
$e$-$p$ collider configurations, after applying the basic acceptance
cuts discussed in the main text.}
\label{tab:xsec-ep}
\end{table}
%=======================================================================

For further signal-background analysis, we have implemented the new
Lagrangian described by Eq.~(\ref{eqn:lag}) in
{\tt Feynrules}\,\cite{Alloul:2013bka} to obtain the {\tt Universal
Feynman Object (UFO)}\,\cite{Degrande:2011ua} files. Then, the UFO model
is used to generate the signal events. The signal and the last three
background events at the parton-level have been generated using
{\tt MadGraph5@aMCNLO}\,\cite{Alwall:2011uj} version 3.4.2 at the lowest
order. At the parton-level event generation, we have imposed basic
`acceptance cuts': ${p_T}_{\mu,e} > 5$~GeV, ${p_T}_j > 10$~GeV,
$|\eta_{e,\mu}| < 3.5$, $|\eta_j| < 8.0$. Here, $p_T$ and $\eta$ denote
the transverse momentum and pseudorapidity, respectively, of the objects
specified by their subscripts. The cross sections corresponding to the
four background processes for four collider configurations are provided
in Table~\ref{tab:xsec-ep}. The parton-level events have then been
showered using {\tt Pythia8}\,\cite{Sjostrand:2014zea} and stored in
{\tt HepMC2}\,\cite{Buckley:2019xhk} formatted files. The
{\tt photoproduction} background events have been generated using
{\tt Pythia8} and stored in {\tt HepMC2} format. These events have then
been passed onto {\tt Delphes}\,\cite{deFavereau:2013fsa} for detector
simulation. For the efficiencies of different objects, we have used the
default card for LHeC and FCC-eh provided in {\tt Delphes} version
3.5.0. We then used {\tt fastjet3}\,\cite{Cacciari:2011ma} to form jets
from the {\tt Delphes} generated Particle Flow (PF) candidates.

\vspace{-16pt}
%=======================================================================
\begin{figure}[h]
\begin{center}
\subfloat[]{\includegraphics[width=0.48\textwidth]{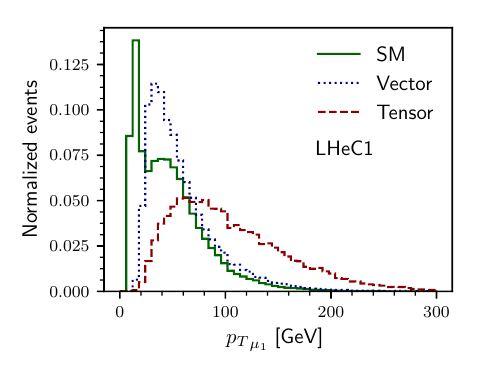}\label{fig:lhec1-ptmu}}
\subfloat[]{\includegraphics[width=0.48\textwidth]{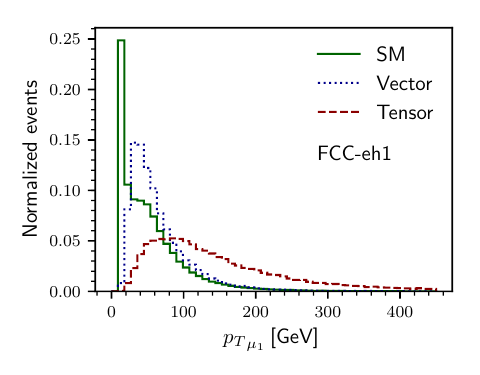}\label{fig:fcc1-ptmu}}\\
\vspace{-12pt}
\subfloat[]{\includegraphics[width=0.48\textwidth]{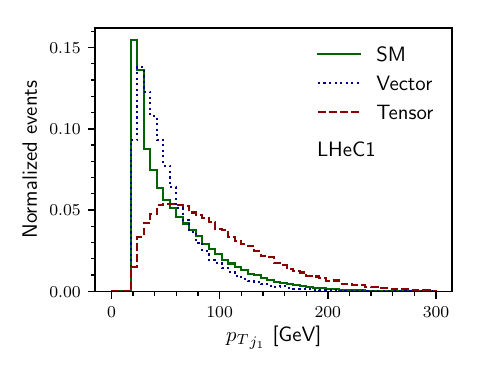}\label{fig:lhec1-ptj}}
\subfloat[]{\includegraphics[width=0.48\textwidth]{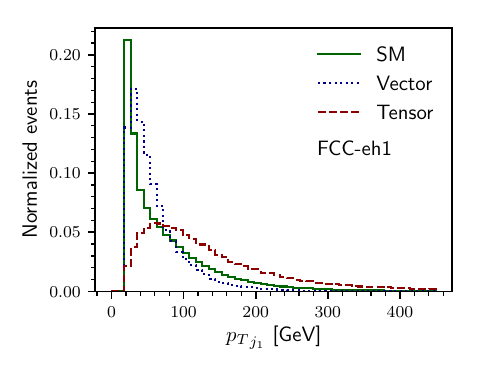}\label{fig:fcc1-ptj}}
\end{center}\vspace{-18pt}
\caption{Normalized distributions of ${p_T}_{\mu_1}$ for both signal and
backgrounds for (a) LHeC1  and (b) FCC-eh1. Normalized distribution of
${p_T}_{j_1}$ for both signal and backgrounds for (c) LHeC1 and (d)
FCC-eh1. For vector-like coupling, we use $g_v=3.68\times 10^{-4}$ and
for tensor-like coupling $g_t=3.00\times 10^{-6}$ GeV$^{-1}$.}
\label{fig:distribution1}
\vspace{-3mm}
\end{figure}
%=======================================================================

%=======================================================================
\begin{figure}[b]
\begin{center}
\subfloat[]{\includegraphics[width=0.48\textwidth]{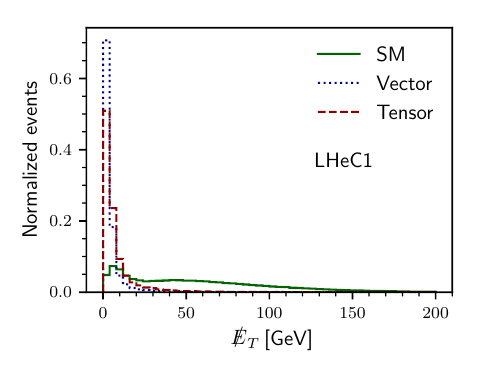}\label{fig:lhec1-misset}}
\subfloat[]{\includegraphics[width=0.48\textwidth]{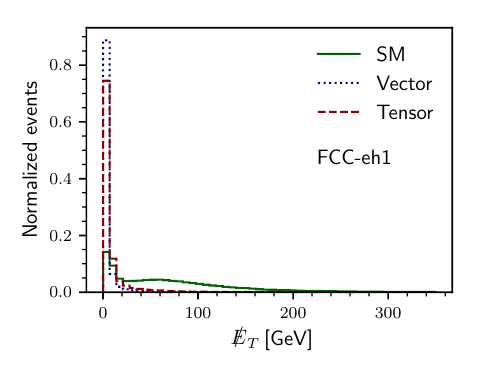}\label{fig:fcc1-misset}}
\vspace{-5mm}
\subfloat[]{\includegraphics[width=0.48\textwidth]{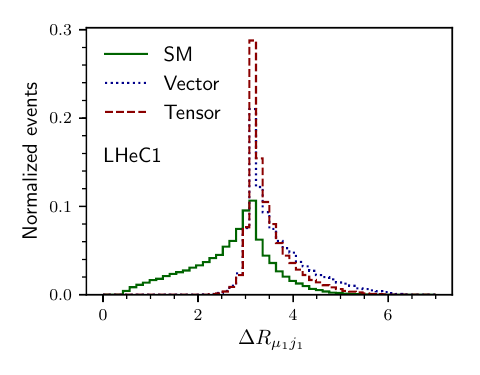}\label{fig:lhec1-drmuj}}
\subfloat[]{\includegraphics[width=0.48\textwidth]{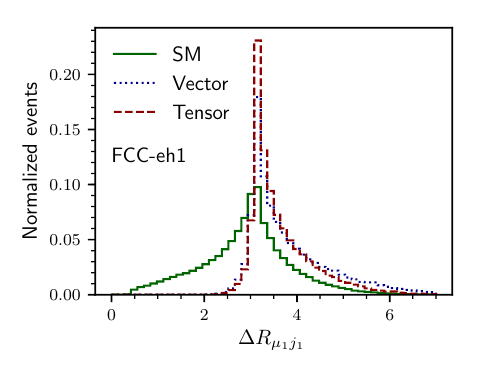}\label{fig:fcc1-drmuj}}
\end{center}\vspace{-5mm}
\caption{Normalized distribution of $\slashed{E}_T$ for both signal and
backgrounds for (a) LHeC1 and (b) FCC-eh1. Normalized distributions are
shown for $\Delta R_{\mu_1 j_{1}}$ for both signal and backgrounds for
(c) LHeC1 and (d) FCC-eh1. The couplings have been taken to be the same
as Fig.~\ref{fig:distribution1}.}
\label{fig:distribution2}
\vspace{-5mm}
\end{figure}
%=======================================================================

The key to any signal-background analysis is to apply some valuable cuts
on some variables, which can give us a signal-favored region with more
signals and fewer backgrounds. To identify those good variables
that discriminate signal and background, we illustrate, in
Figs.~\ref{fig:distribution1} and \ref{fig:distribution2}, the
normalized distributions of some important variables. In
Fig.~\ref{fig:distribution1}, we show the normalized distribution of the
$p_T$ of the leading muon ($\mu_1$) and the leading jet ($j_1$) for the
SM background and the two signal types with non-zero vector and non-zero
tensor coupling. Here, the term `leading' is used to denote the object
that carries the highest $p_T$ among those of its kind. The labeled SM
sample includes all four aforementioned backgrounds, each weighted
appropriately by the product of their cross sections and BRs. The
variable distributions for the signal with non-zero axial-vector and
non-zero axial-tensor couplings are very similar to their non-axial
counterparts and, therefore, are not shown. Figure~\ref{fig:lhec1-ptmu}
corresponds to the distribution of ${p_T}_{\mu_1}$ for LHeC1 and
Fig.~\ref{fig:fcc1-ptmu} corresponds to the same for FCC-eh1 machine.
Figures \ref{fig:lhec1-ptj} and \ref{fig:fcc1-ptj}, show the
distributions of ${p_T}_{j_1}$ for the signal and backgrounds for LHeC1
and FCC-eh1. Because of the momentum dependent factors in tensor
couplings, the distributions of ${p_T}_{\mu_1}$ and ${p_T}_{j_1}$ have
longer tails at higher $p_T$ compared to the vector coupling scenario.
Similarly, in Figs.~\ref{fig:lhec1-misset} and
\ref{fig:fcc1-misset}, we show the distribution of missing transverse
energy $\slashed{E}_T$ for LHeC1 and FCC-eh1. The signal process does
not have any source of $\slashed{E}_T$, so its distribution remains at
lower values. In contrast, all background processes except the
photoproduction process have neutrinos in the final state, leading to a
distribution that extends to higher values. In
Figs.~\ref{fig:lhec1-drmuj} and \ref{fig:fcc1-drmuj}, we illustrate the
distribution of $\Delta R = \sqrt{(\Delta \eta)^2 + (\Delta \phi)^2}$
between the leading muon and the leading jet $\Delta R_{\mu_1 j_1}$ for
LHeC1 and for FCC-eh1. The leading $\mu$ and the leading jet are
back-to-back in the transverse plane for the signal, leading to
$\Delta R$ between them being larger than $\pi$. In contrast, for the
backgrounds, the leading $\mu$ and the leading jet are not necessarily
back-to-back, resulting in a broader distribution towards lower values
of $\Delta R_{\mu_1 j_1}$. These features are reflected in
Figs.~\ref{fig:lhec1-drmuj} and \ref{fig:fcc1-drmuj}.

\vspace{-18pt}
%=======================================================================
\begin{figure}[!h]
\begin{center}
	\subfloat[]{\includegraphics[width=0.48\textwidth]{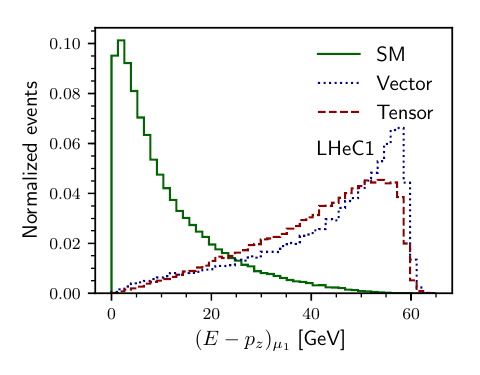}\label{fig:lhec1-E_pz}}
	\subfloat[]{\includegraphics[width=0.48\textwidth]{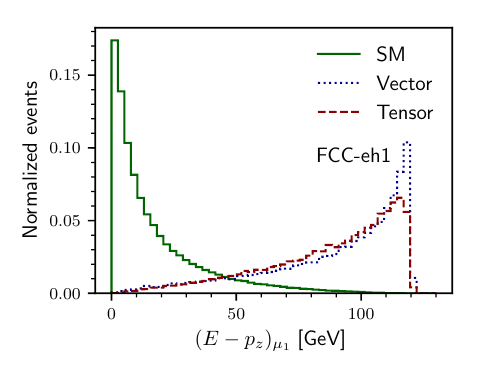}\label{fig:fcc1-E_pz}}
\end{center}\vspace{-5mm}
\caption{Normalized distribution of $(E-p_z)_{\mu_1}$ for both signal
and backgrounds for (a) LHeC1 and (b) FCC-eh1. The same coupling values
as Fig.~\ref{fig:distribution1} have been used.}
\label{fig:distribution3}
\end{figure}\vspace{-8pt}
%=======================================================================

In an electron-proton collider, another useful variable for the signal,
in which no sources of missing energy are expected, is $E-P_z$
variable\,\cite{ZEUS:2005nsy,H1:2011rlk}, defined as the sum of the
differences of energies and $z$-component of the momenta of all final
state visible particles. The $z$-axis is defined in the direction of the
proton beam, and, in the absence of invisible particles in the final
state, $E-P_z$ should be equal to twice the electron beam energy
($E_e$). Since the signal is a $t$-channel process, $(E-p_z)_{\mu_1}$
is expected to be closer to $2E_e$, and $(E-p_z)_{j_1}$ is expected to
be close to zero. In Fig.~\ref{fig:distribution3}, we show the
distribution of $(E-p_z)_{\mu_1}$ for both LHeC1
(Fig.~\ref{fig:lhec1-E_pz}) and FCC-eh1 (Fig.~\ref{fig:fcc1-E_pz}). From
the distribution of $\slashed{E}_T$ and $(E-p_z)_{\mu_1}$, one can
easily guess that these are good variables for reducing the SM
backgrounds over signal with both types of couplings and in both
colliders. Further inspection of the other variables might reveal more
interesting signal regions. Instead of manually inspecting all possible
event variables, we do the signal and background separation with the
help of multivariate analysis. To save on computation time, we apply a
cut of $\slashed{E}_T < 50 \text{GeV}$ before providing the variables to
the multivariate analysis tool.

Keeping these in mind, we have imposed the following preselection cuts
for the signal and the backgrounds.
\begin{eqnarray}
{p_T}_{\mu,e} > 10\ \text{GeV}; \qquad {p_T}_j > 20~\text{GeV}; \qquad
\slashed{E}_T < 50\ \text{GeV};\nonumber \\
|\eta_{e,\mu}| < 3.5; \qquad |\eta_j| < 8.0; \qquad
N_\mu \geq 1; \qquad N_e = 0; \qquad N_j \ge 1.
\label{eqn:preselection}
\end{eqnarray}
After applying the preselection cuts, the products of cross section and
BR, denoted by $\sigma^0$, for our signal and backgrounds for all types
of coupling for different machine energies are listed in
Table~\ref{tab:Xsection}.

%=======================================================================
\begin{table}[h!]
\centering
\begin{tabular}{|c|c|c|c|c|}
\cline{2-5}
\multicolumn{1}{c|}{} 
& \quad {\tt LHeC1}\quad\quad & \quad{\tt LHeC2}\quad\quad& \quad{\tt FCC-eh1 }\quad\quad & \quad{\tt FCC-eh2 }\quad\quad \\  \cline{2-5}	
\multicolumn{1}{c|}{}&\multicolumn{4}{c|}{\tt $\sigma^0$ = Cross section $\times$ BR [fb]} 
\\ \hline
{\tt BKG Photoproduction} & 16.28 & 23.41 & 93.93 & 169.80 \\ \hline	
{\tt BKG mu}              &  4.58 &  9.60 & 20.20 &  31.40 \\ \hline
{\tt BKG e-mu}            &  6.88 & 14.40 & 29.30 &  43.50 \\ \hline
{\tt BKG mu-mu}           &  0.57 &  0.83 &  1.47 &   2.13 \\ \hline
{\tt Vector}              &  0.07 &  0.10 &  0.17 &   0.25 \\ \hline	
{\tt Axial Vector}        &  0.07 &  0.10 &  0.17 &   0.25 \\ \hline
{\tt Tensor }             &  0.36 &  0.65 &  1.27 &   1.95 \\ \hline
{\tt Axial Tensor}        &  0.36 &  0.65 &  1.27 &   1.95 \\ \hline
\end{tabular}
% }
\caption{Cross section times BR of signal and backgrounds with
preselection cuts. For the vector and axial-vector signals, the
coupling constants are
$g_v^{e\mu} = g_{av}^{e\mu} = 3.68\times 10^{-4}$. For the tensor and
axial-tensor signals,
$g_t^{e\mu} = g_{at}^{e\mu} = 5.71\times10^{-6}\text{\,GeV}^{-1}$.}
\label{tab:Xsection}
\end{table}
%=======================================================================

\vspace{-30pt}
%=======================================================================
\subsection{Multivariate analysis} \label{sec:mva}\vspace{-10pt}
After applying the aforementioned cuts on the observables for signal and
background events, we move on to investigate potential improvements in
separating the signal from the background using some established machine
learning methods, such as Gradient Boosted Decision Trees
(BDT)\,\cite{Roe:2004na}. In comparison to a rectangular cut-based
analysis, these techniques have been widely employed in the literature
recently and have been found to offer a superior separation of the
signal from the background. The key objective here is to build a
one-dimensional observable after appropriately combining the important
observables that may effectively distinguish our signals from the
background. We have used {\tt XGBoost}\,\cite{Chen:2016btl} toolkit for
gradient boosting in this analysis.

A total of $15$ input variables, called feature variables, have been
used for the training and validation of our data sample. The variables,
along with their definition and description, are provided in
Table~\ref{featurevar}. With a maximum depth of 3 and a learning rate of
$\sim\!0.001$, we have taken about 1000-3000 estimators for the
gradient-boosted decision tree technique of separation. We have used
70\% of the whole dataset for training purposes and 30\% for validation
in both XGBoost analyses. Overtraining of the BDT is one potential
drawback of this strategy. In cases of overtraining, the test sample
cannot match the training sample's strong discrimination. With our
selection of parameters, we have specifically verified that the
algorithm is not overtrained.

%=======================================================================
\begin{table}[htpb!]
\centering
 \begin{tabular}{||c | c||} 
 \hline
 Variable & Definition \\ [0.5ex] 
 \hline\hline
 ${p_T}_{\mu_1}$ & Transverse momentum of the leading muon \\ 
 ${p_T}_{j_1}$ & Transverse momentum of the leading jet \\
 $\slashed{E}_T$ & Missing transverse energy \\
 $N_{\mu}$ & Number of muons in the event \\
 $N_{j}$ & Number of jets in the event \\
 $m_{\mu_1 j_1}$ & Invariant mass of the leading muon and leading jet \\
 $m_\text{cluster}$ & $\sqrt{m_{j_1}^2 + {p_T}_{j_1}^2} + \sqrt{m_{\mu_1}^2 + {p_T}_{\mu_1}^2} + \slashed{E}_T$
 (The cluster transverse mass\,\cite{Dey:2020tfq})\\
 $m_T$ & $\sqrt{\left(\sqrt{m_{\mu_1}^2 + {p_T}_{\mu_1}^2} + \slashed{E}_T \right)^2 - \left(\vec{p_T}_{\mu_1} + \vec{\slashed{E}}_T\right)^2}
 $
 (Transverse mass) \\
 $H_T$ & Scalar sum of $p_T$'s of all the final state particles \\
 $\Delta \phi_{\mu_1 j_1}$ & Azimuthal angle difference $(\Delta \phi)$ between leading muon and leading jet \\
 $\Delta \phi_{\mu_1 \slashed{E}_T}$ & $\Delta \phi$ between leading muon and missing energy \\
 $\Delta \phi_{j_1 \slashed{E}_T}$ & $\Delta \phi$ between leading jet and missing energy \\
 $\Delta R_{\mu_1 j_1}$ & $\Delta R$ between leading muon and leading jet \\ 
 $(E-p_z)_{\mu_1}$ & Difference between energy and longitudinal momentum of leading muon \\
 $(E-p_z)_{j_1}$ & Difference between energy and longitudinal momentum of leading jet\\ [1ex] 
 \hline
 \end{tabular}
\caption{Feature variables for training in the {\tt XGBoost} toolkit.}
\label{featurevar}
\end{table}\vspace{-16pt}
%=======================================================================

%=======================================================================
\vspace{-16pt}
\section{Result and Discussion}\label{sec:result}\vspace{-10pt}
As mentioned previously, we have considered a single coupling at a time
in order to find out the upper limit on the BR of $Z\to e \mu$ decay.
After a selection criterion, we calculate the signal efficiency
$\epsilon_S$ and background efficiencies $\epsilon_{B,i}$ for
$i^\text{th}$ background with respect to the preselection defined in
Eq.~(\ref{eqn:preselection}). For a given $\epsilon_S$ and
$\epsilon_{B,i}$ for $i^\text{th}$ background, the signal significances
are then calculated as
\begin{eqnarray}\label{eqn:signf}
\text{Signal Significance}\ (\mathfrak{S}) = \sqrt{2 \left[ (S + B) \ln\left(1 + \frac{S}{B} \right) - S \right]}.
\end{eqnarray}
where $S = \sigma_S^0\times\mathcal{L}\times\epsilon_S$, and
$B = \sum_{i}\sigma_{B,i}^0\times\mathcal{L}\times\epsilon_{B,i}$. Here,
$\sigma_S^0$ denotes signal cross section and $\sigma^{0}_{B,i}$
denotes the cross section for the $i^{\rm th}$ background after the
preselection cut, and $\mathcal{L}$ represents the integrated
luminosity. In the last section, we noticed that our vector-only
coupling and axial-vector-only coupling are not noticeably different
from each other. The same is true for tensor-only and axial-tensor-only
couplings. Therefore, the results will be discussed jointly for vector
and axial-vector couplings, and for tensor and axial-tensor couplings,
in the following subsections.

%=======================================================================
\vspace{-16pt}
\subsection{Vector and axial-vector coupling}
\begin{figure}[!h]
\begin{center}
\subfloat[]{\includegraphics[width=0.49\textwidth]{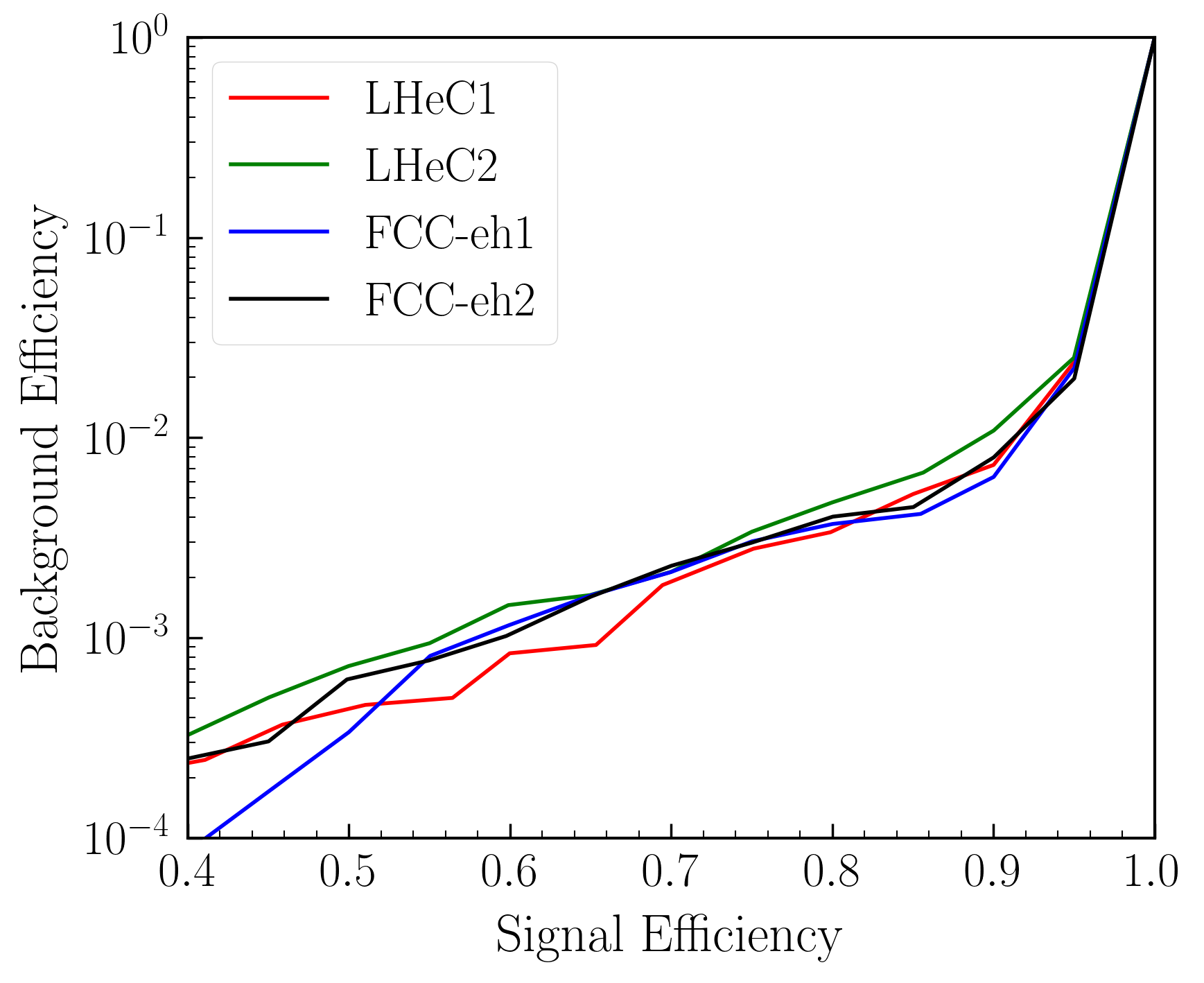}\label{fig:ROC-vec}}
\subfloat[]{\includegraphics[width=0.465\textwidth]{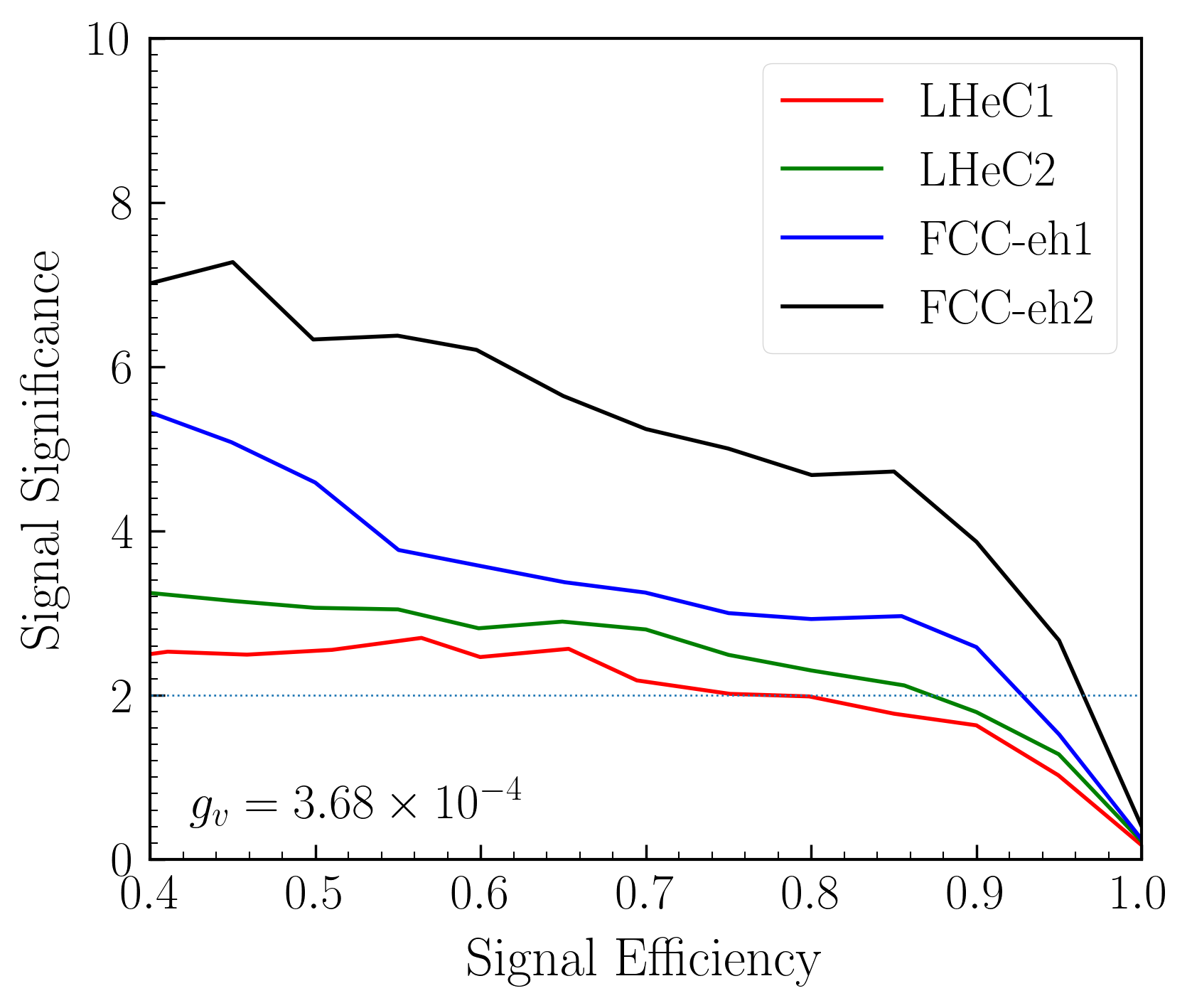}\label{fig:signi-vec}}
\end{center}\vspace{-5mm}
\caption{(a) ROC curves on the BDT classifier for signal vs.~background.
(b) Variations of signal significance as a function of signal efficiency
of the  BDT classifier with $g_v=3.68\times 10^{-4}$. The significances
are calculated for integrated luminosities of 50 fb$^{-1}$ for LHeC1, 90
fb$^{-1}$ for LHeC2, 80 fb$^{-1}$ for FCC-eh1, and 150 fb$^{-1}$ for
FCC-eh2. The horizontal dotted line indicates 2$\sigma$ signal
significance. In both panels, only the vector coupling has been
considered.}
\label{fig:ROCsigni-vec}
\end{figure}

\vspace{-10pt}
For the discussion in this subsection, we have chosen only
$g_v = 3.68 \times 10^{-4}$ to be non-zero. We first show, in
Fig.~\ref{fig:ROC-vec}, the performance of the BDT network through the
Receiver Operating Characteristic (ROC) curve. The good performance of
the network is evident from the figure. The background acceptance is
below 0.001 with 50\% signal efficiency. This high performance is
because most of the backgrounds contain neutrinos, a source of missing
energy, in contrast to the signal, where there is no source of missing
energy.
%=======================================================================
%=======================================================================
The network returns the BDT classifier, which can then, in principle, be
used to set a cut and calculate the signal significance. We have shown
these signal significances as a function of signal efficiencies.

The variation of signal significance as a function of signal efficiency
is shown in Fig.~\ref{fig:signi-vec} for all four machine energies as
described in Table~\ref{tab:coll-param}. A feature of the curves shows
that the signal significance is the best in the region 40\% to 70\%
signal efficiencies. The background efficiencies in this region are
small, varying between $10^{-4}$ to $10^{-2}$. The region below 40\% of
signal efficiency features a low number of background events (sometimes
zero). As a result, this area is not reliable as there is a mismatch
between the train and test samples.
%=======================================================================
\vspace{-16pt}
\begin{figure}[!h]
\begin{center}
\subfloat[]{\includegraphics[width=0.47\textwidth]{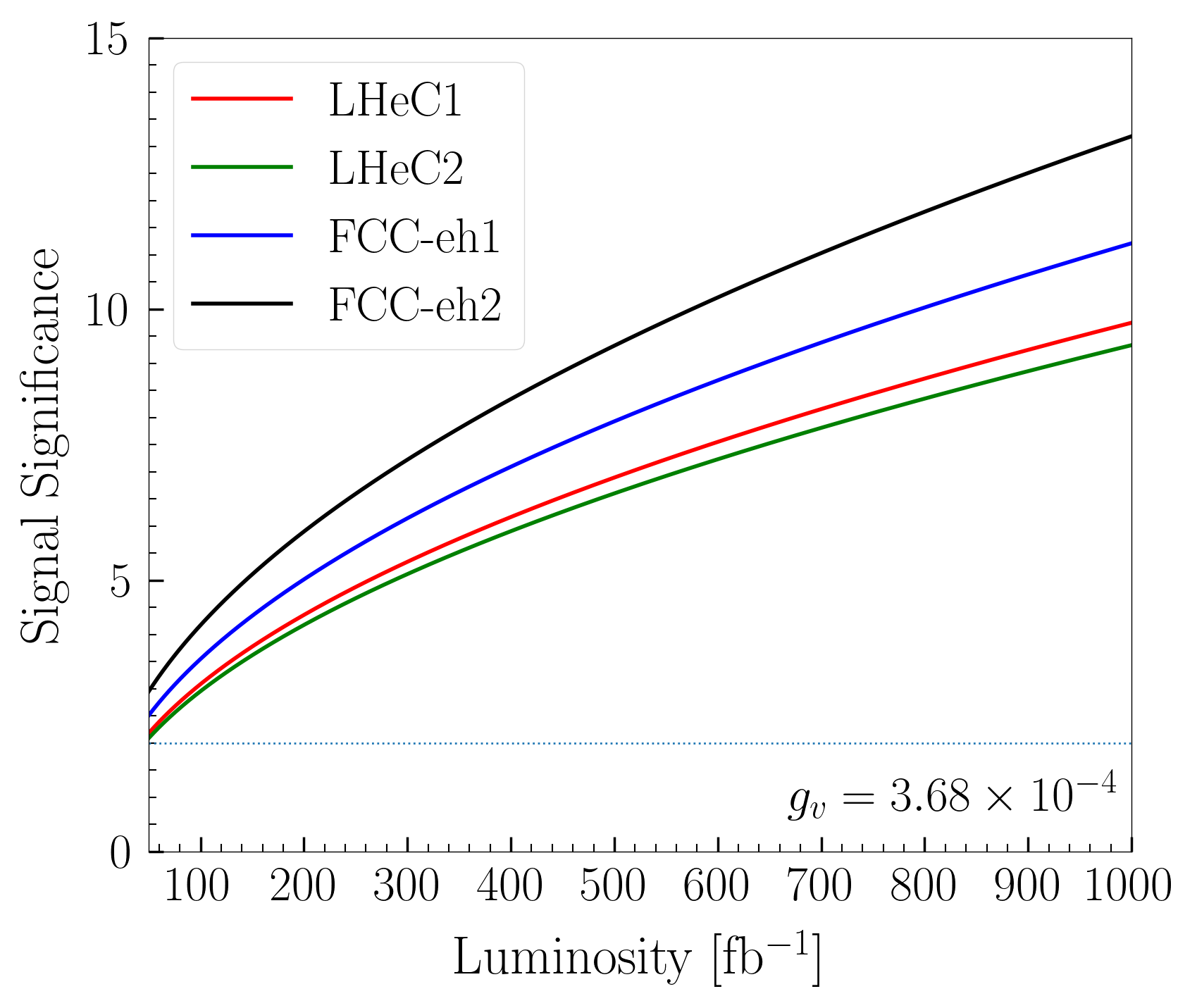}\label{fig:signi-lumi-vec}}
\subfloat[]{\includegraphics[width=0.49\textwidth]{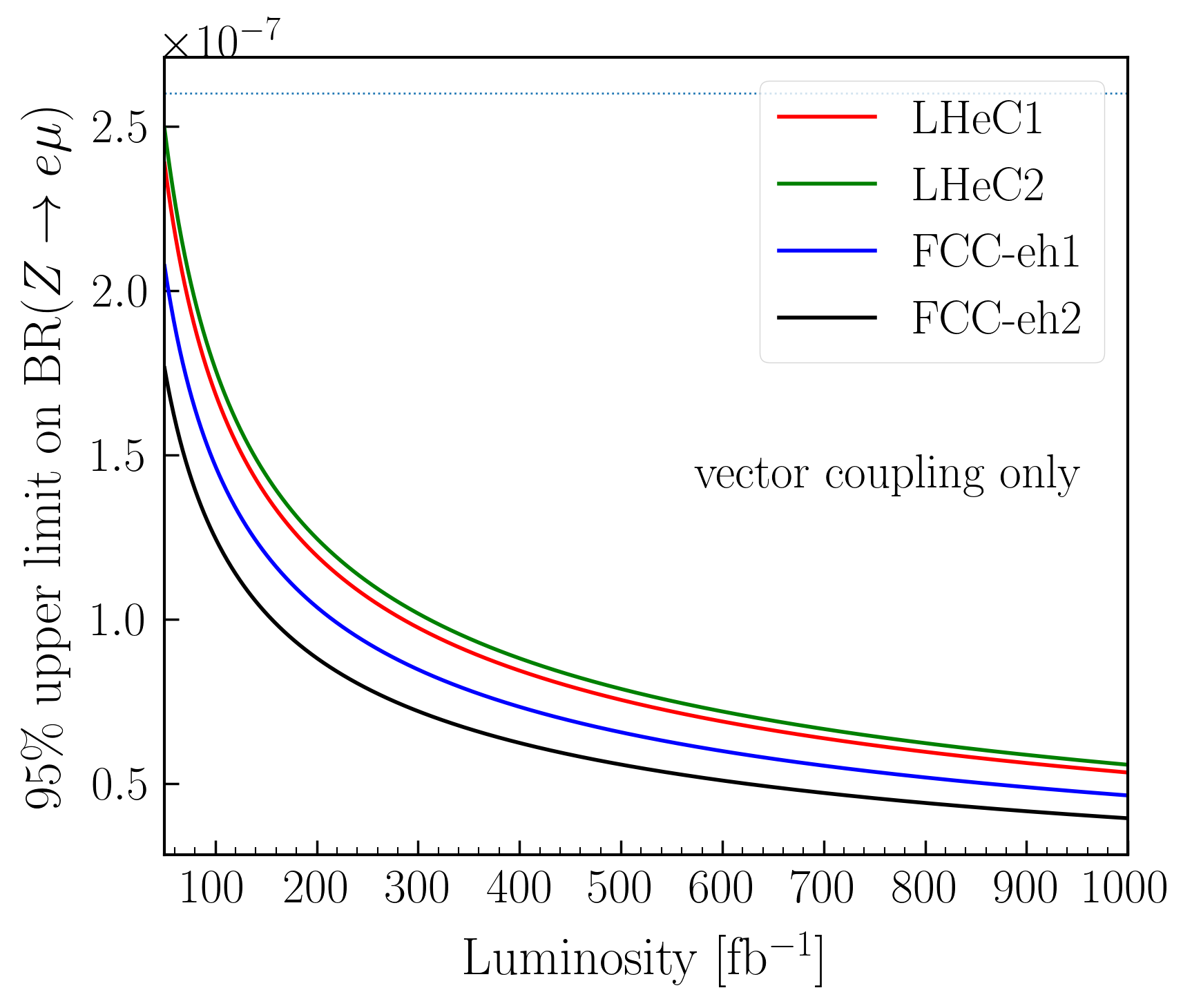}\label{fig:bound-vec}} \\
\end{center}
\vspace{-18pt}
\caption{(a) Signal significance as a function of integrated luminosity;
the horizontal dotted line represents the 2$\sigma$ signal significance.
(b) Variations of the 95\% C.L.~upper limit on BR($Z\to e\mu$) as a
function of integrated luminosity; the horizontal dotted line indicates
the current ATLAS upper limit on BR($Z\to e\mu$). In both panels, only
the vector coupling has been considered, and the signal efficiency is
taken to be 70\% for all collider setups.}
\label{fig:signibound-vec}
\end{figure}\vspace{-10pt}
%=======================================================================

We then choose our working points to be $\epsilon_S=0.7$. We also
ensured that there was a good agreement in the distribution between the
test and train samples at those working points. The variation of signal
significance as a function of integrated luminosity has been shown in
Fig.~\ref{fig:signi-lumi-vec} for four different colliders. One can
observe that the signal significance is distinctively better
($\sim\!2.0 - 3.0$ times better) in FCC-eh1 and FCC-eh2 compared to all
LHeC energies. Furthermore, for a given luminosity, the signal
significance varies as the square of the coupling constant $g_v$. Now,
we can, in principle, put an upper limit on the coupling constant at a
$2\sigma$ signal significance. That is, to say, the value of $g_v$ at
which 2$\sigma$ signal significance is achieved. These limits on $g_v$
can then be translated into the limit on the LFV $Z$ boson BRs by using
Eq.~(\ref{eqn:brZ}). We plot, in Fig.~\ref{fig:bound-vec}, the upper
limit on BR($Z\to e \mu$) at 95\% C.L.~as a function of luminosity for
four different machines. In Fig.~\ref{fig:bound-vec}, the horizontal
line represents the current bound by the ATLAS
collaboration\,\cite{ATLAS:2022uhq}, which is achievable below 50
fb$^{-1}$ in all the future electron-proton colliders.

There are a few points to note from Figs.~\ref{fig:ROCsigni-vec} and
\ref{fig:signibound-vec}. Firstly, the ROC curves in
Fig.~\ref{fig:ROC-vec} compare the performances in terms of the signal
and background efficiencies, defined as the fraction of signal or
background events surviving after applying cuts on the BDT classifier.
The ROC curves indicate that, for the same signal efficiency, the
background rejection is better in LHeC1 than in LHeC2, implying that
LHeC1 performs better. Furthermore, as seen from
Table~\ref{tab:Xsection}, the total background cross section increases
by about 70\% from LHeC1 to LHeC2, while the corresponding increase for
the signal is only around 40\% for the vector and axial-vector coupling
scenarios. These two disadvantages of LHeC2 with respect to LHeC1 --
one arising from the ROC performances and the other from how the signal
and background cross sections scale with center-of-mass energy -- can be
realized in Figs.~\ref{fig:signi-lumi-vec} and \ref{fig:bound-vec} by
comparing sensitivities for a given integrated luminosity. As a result,
for the same integrated luminosity, LHeC2 is less sensitive than LHeC1,
even though LHeC2 has a higher energy reach. The relatively better
significance in LHeC2 compared to LHeC1 in Fig.~\ref{fig:signi-vec} is
primarily due to the higher integrated luminosity of LHeC2. However,
between FCC-eh1 and FCC-eh2, the latter performs better owing to the
similar sensitivities in the ROC curves and a somewhat larger signal
increase of about 50\%.

\vspace{-20pt}
\subsection{Tensor and axial-tensor coupling}
\begin{figure}[!h]
\begin{center}
\subfloat[]{\includegraphics[width=0.49\textwidth]{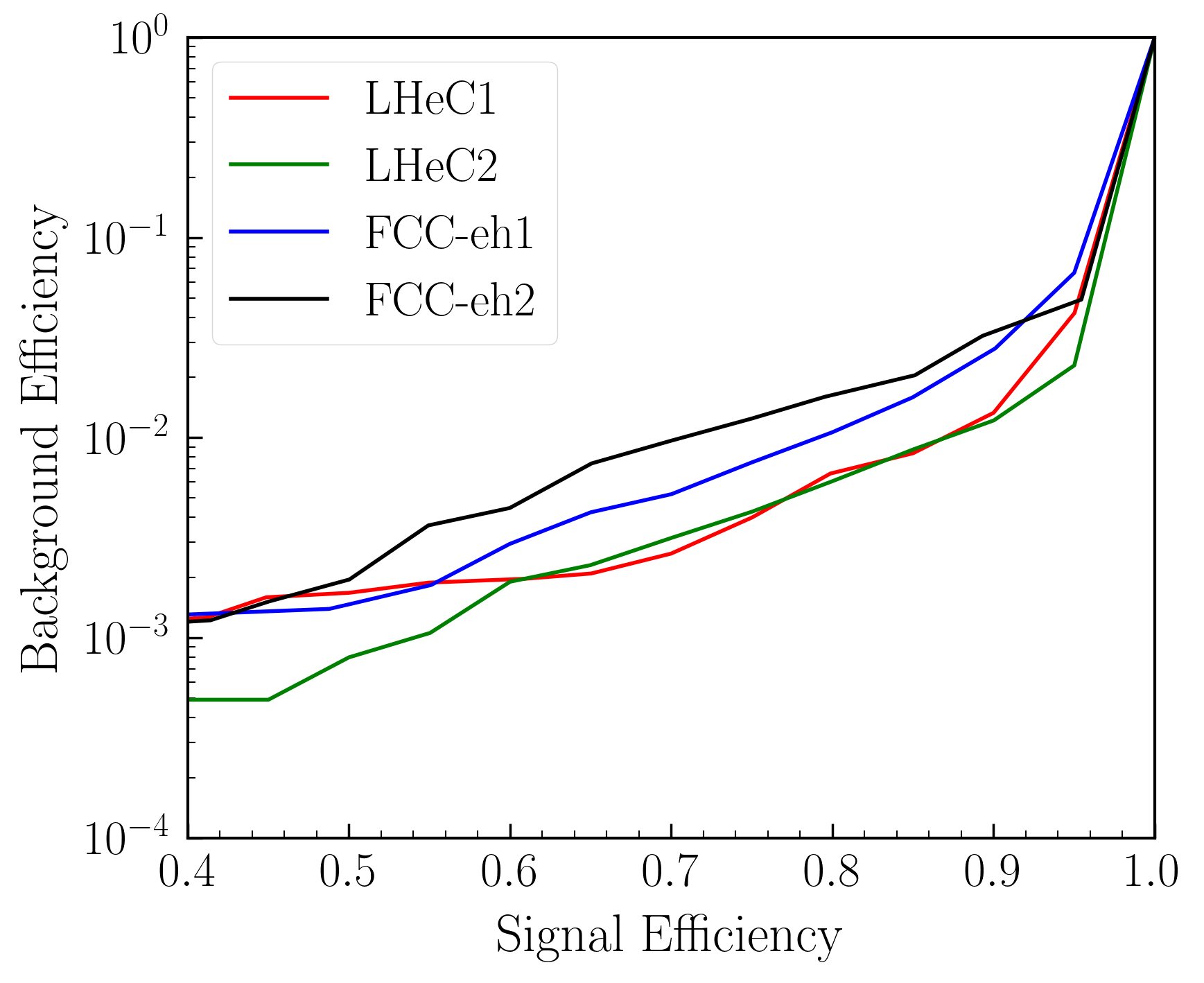}\label{fig:ROC-ten}}
\subfloat[]{\includegraphics[width=0.47\textwidth]{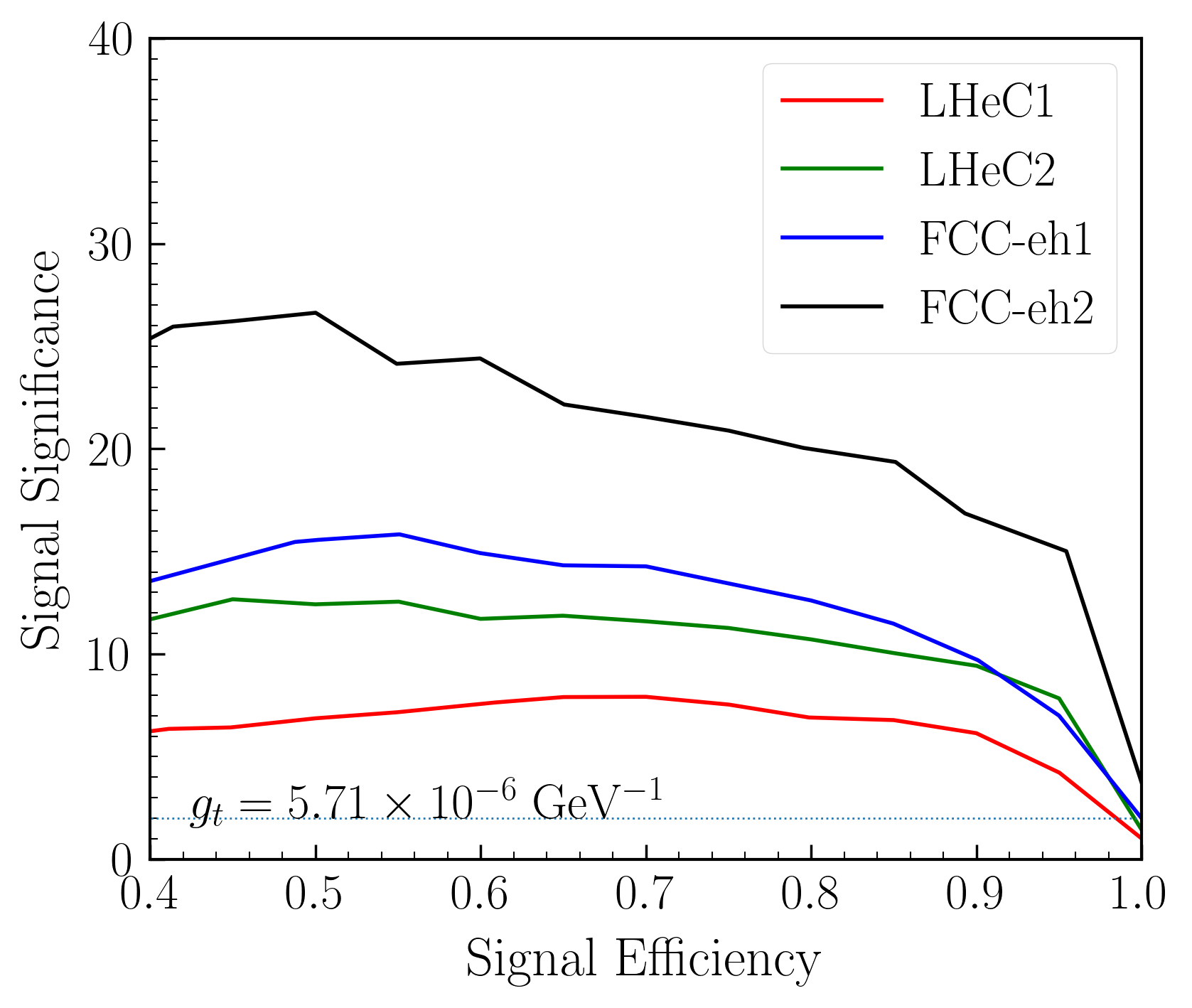}\label{fig:signi-ten}}
\end{center}\vspace{-16pt}
\caption{(a) ROC curves on the BDT classifier for signal vs.~background.
(b) Variations of signal significance as a function of signal efficiency
of the  BDT classifier with $g_t=5.71\times 10^{-6}$~GeV$^{-1}$. The
significances are calculated for integrated luminosities of 50 fb$^{-1}$
for LHeC1, 90 fb$^{-1}$ for LHeC2, 80 fb$^{-1}$ for FCC-eh1, and 150
fb$^{-1}$ for FCC-eh2. The horizontal dotted line indicates to 2$\sigma$
signal significance. In both panels, only the tensor coupling has been
considered.}
\label{fig:ROCsigni-ten}
\end{figure}

\vspace{-12pt}
Similar to the previous subsection, we then performed an analysis with
the tensor LFV coupling. In this case, we have set only the tensor
coupling to be non-zero at $g_t = 5.71\times10^{-6}~\text{GeV}^{-1}$. We
first show the ROC curves for four different machines in
Fig.~\ref{fig:ROC-ten}.
%=======================================================================
%=======================================================================
The ROC curves actually feature a very good separation between the
signal and the backgrounds. In all the machine energies, the background
acceptance rate is between $10^{-3}$ to $10^{-2}$ for signal
efficiencies between 40\% to 80\%. Signal significance as a function of
signal efficiency has been plotted in Fig.~\ref{fig:signi-ten}. We can
see that, in the case of tensor-only coupling, FCC-eh runs can provide
better (approximately a factor of 1.2 -- 3.0) significance than LHeC
runs. Although a quick comparison between Fig.~\ref{fig:signi-vec} and
Fig.~\ref{fig:signi-ten} may indicate that the tensor-only coupling
provides better signal significances, we note that they should not be
compared as the coupling values, at this point, are arbitrary and they
are for demonstration only.

%=======================================================================
%\vspace{-10pt}
\begin{figure}[!h]
\begin{center}
\subfloat[]{\includegraphics[width=0.47\textwidth]{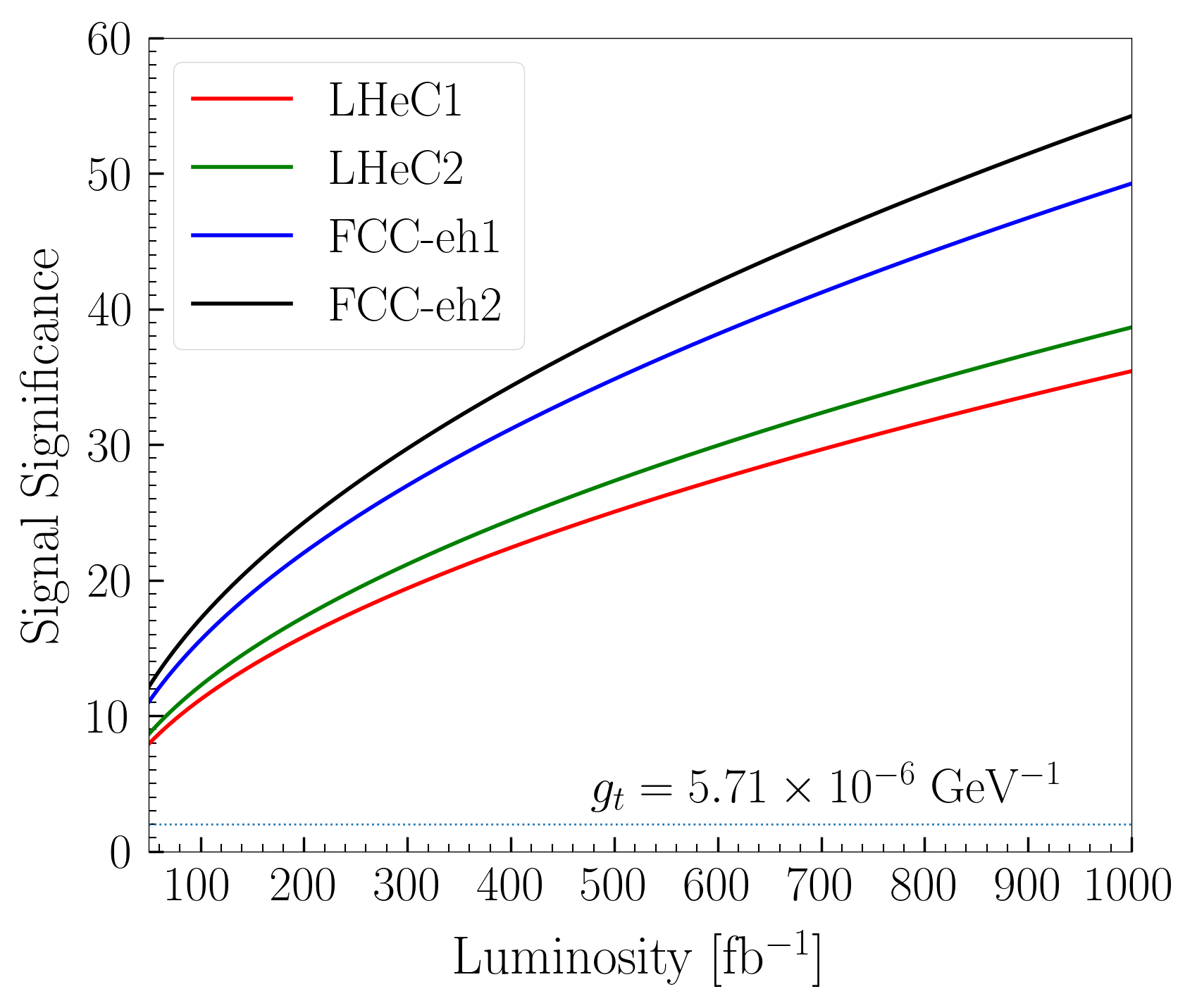}\label{fig:signi-lumi-ten}}
\subfloat[]{\includegraphics[width=0.49\textwidth]{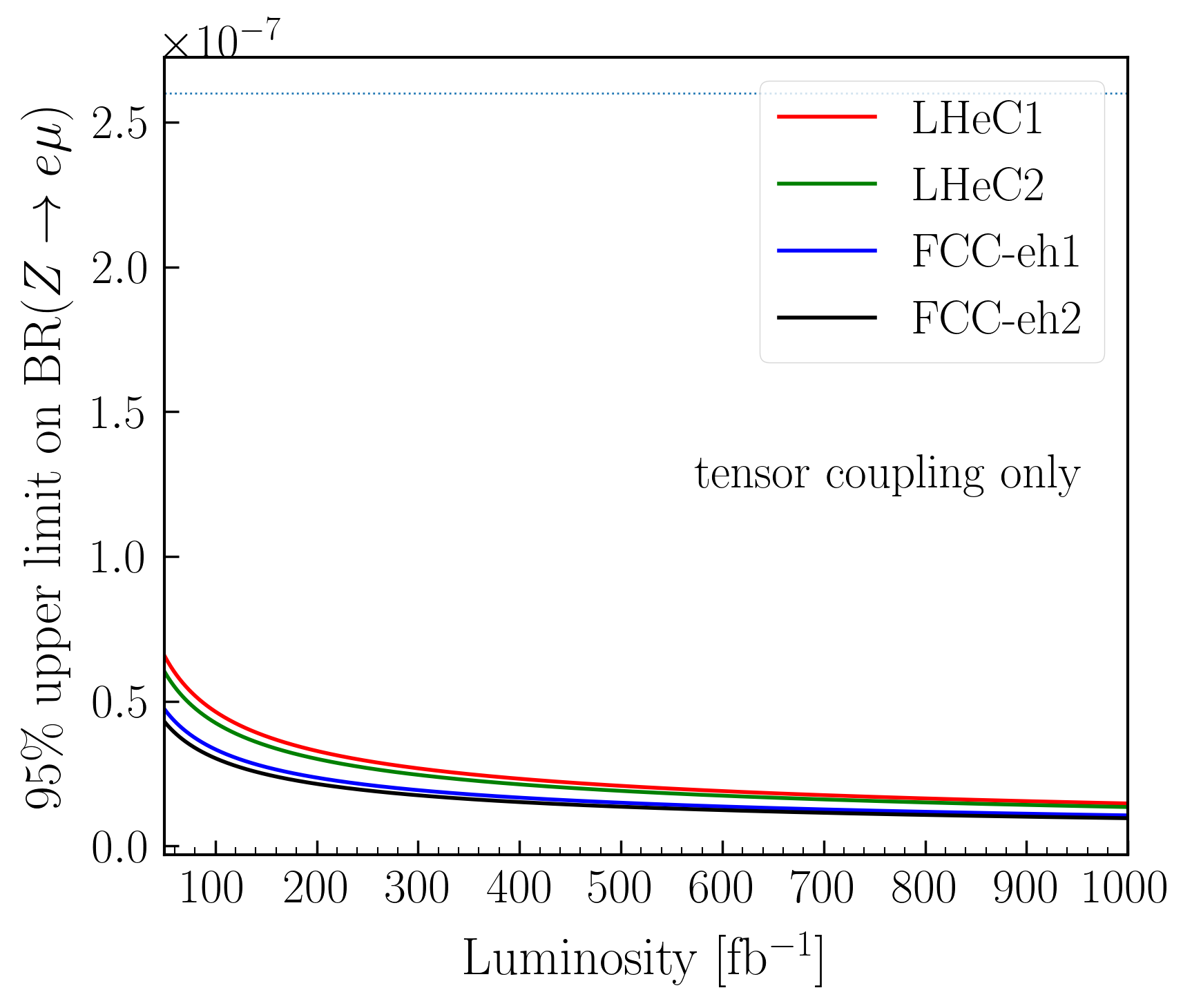}\label{fig:bound-ten}} \\
\end{center}\vspace{-10pt}
\caption{(a) Signal significance as a function of integrated luminosity;
the horizontal dotted line represents the 2$\sigma$ signal significance.
(b) Variations of the 95\% C.L.~upper limit on BR($Z\to e\mu$) as a
function of integrated luminosity; the horizontal dotted line indicates
the current ATLAS upper limit on BR($Z\to e\mu$). In both panels, only
the tensor coupling has been considered, and the signal efficiency is
taken to be 70\% for all collider setups.}
\label{fig:signibound-ten}
\end{figure}
%=======================================================================
We then choose our working points to be $\epsilon_S=0.7$. We show in
Fig.~\ref{fig:signi-lumi-ten} the variation of signal significance as a
function of luminosity. To avoid overtraining, we ensure good agreement
in the distribution between the test and train samples in our chosen
$\epsilon_S$. The signal significances are above 2$\sigma$ even with 50
fb$^{-1}$ in the case of LHeC runs and are well above 2$\sigma$ in the
case of FCC-eh runs.
The projected upper limit on BR($Z\to e \mu$) as a function of
integrated luminosity at $\epsilon_S=0.7$ has been plotted in
Fig.~\ref{fig:bound-ten}. In the case of tensor-only coupling, one can
see that the current bound on BR($Z\to e \mu$) at 95\% C.L.~can easily
be achieved with less than 50 fb$^{-1}$ integrated luminosity in all
four machine energies.

Here, unlike the vector-only coupling scenario, LHeC2 performs better
compared to LHeC1. The primary reason is the significant increase in
signal cross section, which is about 80\%. A similar feature is observed
between FCC-eh1 and FCC-eh2.

\vspace{-16pt}
\subsection{Bounds on LFV branching ratio}\vspace{-10pt}
We now put together the results from the last two subsections by
providing projected bounds on BR($Z\to e\mu$) for four different machine
energies. As we mentioned previously, we consider only one effective
coupling at a time. These projected bounds on the BR and the effective
coupling are shown in Tables~\ref{tab:lhec} and \ref{tab:fcc}. We have
considered a one-year runtime for each machine and the integrated
luminosity as provided in Table~\ref{tab:coll-param}.

%=======================================================================
\begin{table}[h!]
\centering
\resizebox{\textwidth}{!}{
\begin{tabular}{|c|c|c|c|c|c|c|c|c|}
\cline{2-9}
\multicolumn{1}{c|}{} 
& \multicolumn{8}{c|}{\tt Projected upper bound at 95\% C.L.~on BR($Z \rightarrow e \mu$) and couplings} \\ [-3pt]
\multicolumn{1}{c|}{} 
& \multicolumn{8}{c|}{\tt along with number of signal and background events} \\  
\cline{2-9}
\multicolumn{1}{c|}{} 
& \multicolumn{4}{c|}{\tt LHeC1}& \multicolumn{4}{c|}{\tt LHeC2} \\ \hline
{\tt Coupling Type}  & {\tt\,BR($Z \rightarrow e \mu$)}   & ~~{\tt $g_{v/t}$}~~ & ~~S~~ & ~~B~~ &  {\,\tt BR($Z \rightarrow e \mu$)} &  ~~{\tt $g_{v/t}$}~~ & ~~S~~ & ~~B~~ \\ \hline		
{\tt Vector}    & $2.36\times 10^{-7}$ & $3.50\times 10^{-4}$ & $2.9\pm1.7$ & $1.3\pm1.1$ & $1.79\times 10^{-7}$ & $3.04\times 10^{-4}$ & $5.3\pm2.3$ & $5.5\pm2.3$ \\ \hline
{\tt Axial Vector} & $2.11\times 10^{-7}$ & $3.30\times 10^{-4}$ & $2.6\pm1.6$ & $1.0\pm1.0$ & $1.13\times 10^{-7}$ & $2.41\times 10^{-4}$ & $3.3\pm1.8$ & $1.9\pm1.4$ \\ \hline
{\tt Tensor } & $4.60\times 10^{-8}$ & $2.39\times 10^{-6}$ GeV$^{-1}$ & $3.3\pm1.8$ & $1.8\pm1.4$ & $3.23\times 10^{-8}$ & $2.00\times 10^{-6}$ GeV$^{-1}$ & $6.3\pm2.5$ & $8.1\pm2.8$\\  \hline
{\tt Axial Tensor} & $5.09\times 10^{-8}$ & $2.52\times 10^{-6}$ GeV$^{-1}$ & $3.7\pm1.9$ & $2.3\pm1.5$ & $2.34\times 10^{-8}$ & $1.71\times 10^{-6}$ GeV$^{-1}$ & $4.6\pm2.1$ & $3.9\pm2.0$\\  \hline
\end{tabular}
}
\caption{Projected upper bound at 95\% C.L.~on BR($Z \rightarrow e \mu$)
and LFV couplings for the LHeC collider. Current upper bound at 95\%
C.L.~on BR($Z \rightarrow e \mu$) from ATLAS is
$2.62 \times 10^{-7}$\,\cite{ATLAS:2022uhq}.}
\label{tab:lhec}
\end{table}
%=======================================================================

\vspace{-8pt}
%=======================================================================
\begin{table}[h!]
\centering
\resizebox{\textwidth}{!}{
\begin{tabular}{|c|c|c|c|c|c|c|c|c|}
\cline{2-9}
\multicolumn{1}{c|}{} 
& \multicolumn{8}{c|}{\tt Projected upper bound at 95\% C.L.~on BR($Z \rightarrow e \mu$) and couplings} \\  [-3pt]
\multicolumn{1}{c|}{} 
& \multicolumn{8}{c|}{\tt along with number of signal and background events} \\  
\cline{2-9}
\multicolumn{1}{c|}{} 
& \multicolumn{4}{c|}{\tt FCC-eh1 }& \multicolumn{4}{c|}{\tt FCC-eh2 } \\ \hline
{\tt Coupling Type}  &  {\tt\,BR($Z \rightarrow e \mu$)} & ~~{\tt $g_{v/t}$}~~ & ~~S~~ & ~~B~~ & {\tt\,BR($Z \rightarrow e \mu$)} &  ~~{\tt $g_{v/t}$}~~  & ~~S~~ & ~~B~~\\ \hline		
{\tt Vector}       & $1.54\times 10^{-7}$ & $2.82\times 10^{-4}$ & $7.9\pm2.8$ & $13.1\pm3.6$ & $9.32\times 10^{-8}$ & $2.19\times 10^{-4}$ & $13.7\pm3.7$ & $42.9\pm6.5$\\ \hline
{\tt Axial Vector} & $1.75\times 10^{-7}$ & $3.00\times 10^{-4}$ & $9.0\pm3.0$ & $17.3\pm4.2$ & $4.64\times 10^{-8}$ & $1.55\times 10^{-4}$ & $6.8\pm2.6$ & $9.6\pm3.1$ \\ \hline
{\tt Tensor }      & $2.85\times 10^{-8}$ & $1.88\times 10^{-6}$ GeV$^{-1}$ & $12.0\pm3.5$ & $32.2\pm5.7$ & $2.01\times 10^{-8}$ & $1.58\times 10^{-6}$ GeV$^{-1}$ & $27.5\pm5.2$ & $180.8\pm13.4$\\  \hline
{\tt Axial Tensor} & $2.85\times 10^{-8}$ & $1.88\times 10^{-6}$ GeV$^{-1}$ & $12.0\pm3.5$ & $32.4\pm5.7$ & $5.02\times 10^{-9}$ & $7.90\times 10^{-7}$ GeV$^{-1}$ & $6.9\pm2.6$ & $9.8\pm2.1$\\  \hline
\end{tabular}
}
\caption{Projected upper bounds at 95\% C.L.~on
BR($Z \rightarrow e \mu$) and LFV couplings for FCC-eh collider. Current
upper bound at 95\% C.L.~on BR($Z \rightarrow e \mu$) from ATLAS is
$2.62\times 10^{-7}$\,\cite{ATLAS:2022uhq}.}
\label{tab:fcc}
\end{table}\vspace{-10pt}
%=======================================================================

As was hinted at by the discussions in the previous two subsections, we
note that the future electron-proton collider has the potential to
provide a stronger upper bound. As mentioned in
Table~\ref{tab:coll-param}, if we look in the LHeC case, for run 1, an
integrated luminosity around $50$ fb$^{-1}$ can be achieved, and for run
2, it can be $\sim\!90$ fb$^{-1}$ after collecting one year of data.
Considering that we can marginally improve the limit on
BR($Z \rightarrow e \mu$) compared to the current bound provided by the
ATLAS collaboration\,\cite{ATLAS:2022uhq} for the case of vector-only
and axial-vector-only coupling. On the other hand, if the new physics
LFV couplings are governed by tensor or axial-tensor coupling, the bound
on BR($Z\to e\mu$) can be $\sim\!5$ times stronger than the current
upper limit. Overall, there is a scope for improvement in the current
bounds at both runs of the LHeC machine. These numbers are calculated
assuming that the run is for one year only. With more than one year of
run, there is further scope for improvement.

Let us look at the FCC-eh case now. In that case, we can get an
integrated luminosity of 80 fb$^{-1}$ for run-1 and it can be 150
fb$^{-1}$ for run-2 after collecting all data for one year. So, if one
focuses only on the luminosity after one year run for FCC-eh1, the
improvement on the BR($Z \rightarrow e \mu$) can be $\sim\!2$ times
stronger than the existing bound set by ATLAS for the case of
vector-only and axial-vector-only coupling. Whereas, for tensor-only or
axial-tensor-only coupling, the improvement can be more than a factor of
10. On the other hand, a more remarkable improvement can be shown if
we take a look at the FCC-eh2 case. In that case, after the collection
of one-year data, the bound can be $\sim\!9$ times stronger compared to
the existing bound. Moreover, with longer runtimes at future $e$-$p$
colliders, there is potential to achieve even better sensitivity.

We also note that there are projections for LFVZD at the Future Circular
electron-positron Collider (FCC-ee)\,\cite{Dam:2018rfz}, which estimate
an upper limit on $\text{BR}(Z \to e\mu)$ at the level of
$10^{-8}$--$10^{-10}$. Our projected sensitivity lies in the range of
$10^{-8}$--$10^{-9}$. The approach in Ref.\,\cite{Dam:2018rfz} is
similar to that of the ATLAS analysis but applied to FCC-ee projections.
The probe of BR($Z \to e\mu$) in that work is based on a search for
oppositely charged $e\mu$ pairs around the $Z$ mass. They considered two
main backgrounds: $Z \to \tau\tau$ (leptonic decays) and $\mu \to e$
misidentification in $Z\to \mu\mu$ decays. The latter gives the dominant
contribution, leading to an upper limit of about $10^{-8}$--$10^{-10}$
on BR($Z \to e\mu$). Their estimate, however, relies on approximate
assumptions about conversion probability and not on a detailed
signal-background analysis. Our approach is different, as we study the
$\mu+j$ final state from a flavor-violating vertex based on a detailed
signal-background analysis.

%=======================================================================
\vspace{-20pt}
\section{Summary and Conclusion} \label{sec:summary}\vspace{-12pt}
The search for LFV is an important area of research since experimental
observation of LFV will hint towards new physics beyond the SM. We
performed an analysis of such an LFV scenario in the context of future
electron-proton colliders. We focused on the LFV coupling of $Z$ boson
to an electron-muon pair and on the search for it in the channel with a
single $\mu$ plus an associated $j$. This $\mu+j$ final state without
any large missing $E_T$ can only come from LFV scenarios for which the
SM background is very small. If no signal is found at the collider, an
upper limit on the LFV coupling of the $Z$ boson can be set. The upper
limit on the coupling can then be translated to the BR of $Z\to e\mu$
decay. For this work, we have considered one out of four different types
of coupling at a time, namely vector, axial-vector, tensor, and
axial-tensor coupling.

We carried out our calculations for two future electron-hadron
colliders, namely LHeC and FCC-eh. Detector effects have been simulated
with {\tt Delphes} version 3.5.0 and the default detector cards for the
respective colliders. A detailed estimation of systematic uncertainties
or detector effects, such as beam conditions and data quality
variations, is beyond the scope of the present work. The projected
sensitivities should therefore be viewed as indicative estimates. While
a full experimental analysis may lead to quantitative differences at the
level of factors of order unity, the qualitative conclusions of this
work are expected to remain valid.

We have used a multivariate technique to maximize the discovery
potential of such LFV signals in the presence of SM backgrounds. The
absence of discovery at 2$\sigma$ has then been translated to the upper
limit on the LFV Z couplings and on the BR($Z\to e\mu$). We observed
that vector-only coupling and axial-vector-only coupling provide
approximately similar sensitivity. The same is true for tensor-only and
axial-tensor-only couplings.

For LHeC run-1, with 50 fb$^{-1}$ integrated luminosity after one year
run, an upper limit of up to $2.11 \times 10^{-7}$ at 95\% C.L.~on the
BR($Z \rightarrow e \mu$) can be set if the LFV coupling is completely
either vector or axial-vector coupling. This bound marginally improves
over the existing bound, which is set as $2.62\times 10^{-7}$ by ATLAS,
on such a BR. If we consider that either the tensor or axial-tensor
coupling is responsible for the lepton flavor violation, the projected
bounds can be made stronger. For tensor coupling, our projection is
BR$(Z \rightarrow e \mu)< 4.60 \times 10^{-8}$ at 95\% C.L. For LHeC
run-2, with 90 fb$^{-1}$ integrated luminosity after one year run, the
projected 95\% C.L.~limit becomes
BR$(Z \rightarrow e \mu) < 1.79 \times 10^{-7}$ for vector-only
coupling scenario and BR$(Z\rightarrow e \mu) < 3.23 \times 10^{-8}$
for tensor-only coupling scenario.

A significant improvement can be made at the FCC-eh machine. In that
case, for run-1 with 80 fb$^{-1}$ luminosity after one year run, the
estimated projection becomes
BR$(Z \rightarrow e \mu) < 1.54 \times 10^{-7}$ with vector-only
coupling and BR$(Z \rightarrow e \mu) < 2.85 \times 10^{-8}$ with
tensor-only coupling. The best case scenario happens for FCC-eh run-2
which will collect data with an integrated luminosity of 150~fb$^{-1}$
after one year of running. The projected bounds in that case are
BR$(Z \rightarrow e \mu) < 9.32 \times 10^{-8}$ and
BR$(Z \rightarrow e \mu) < 2.01 \times 10^{-8}$ with vector-only and
tensor-only couplings, respectively. With more years of running for
both the machines, the projection on BR($Z\to e\mu$) is expected to get
better.

%=======================================================================
\section*{Data availability}\vspace{-12pt}
This study is based on simulated data generated using standard HEP
software packages. All details necessary to reproduce the simulations
and analyses are provided within the article.

\vspace{-16pt}
%=======================================================================
\section*{Acknowledgements}\vspace{-12pt}
The authors thank the 2022 November Meeting at IISER-K for providing an
environment for fruitful discussions. The authors acknowledge the
support of the Kepler Computing facility, maintained by the Department
of Physical Sciences, IISER Kolkata, and the RECAPP cluster facility for
various computational needs. A.K.B.~acknowledges the support from the
Department of Atomic Energy, Government of India, for the Regional
Centre for Accelerator-based Particle Physics~(RECAPP).
A.D.~acknowledges financial support from Science Foundation Ireland
Grant 21/PATH-S/9475 (MOREHIGGS) under the SFI-IRC Pathway Programme.

%\bibliographystyle{JHEP}
%\bibliography{refs}

\begin{thebibliography}{10}

\bibitem{SNO:2002tuh}
{\scshape SNO} collaboration, \emph{{Direct evidence for neutrino flavor
  transformation from neutral-current interactions in the Sudbury Neutrino
  Observatory}},
  \href{https://doi.org/10.1103/PhysRevLett.89.011301}{\emph{Phys. Rev. Lett.}
  {\bfseries 89} (2002) 011301}
  [\href{https://arxiv.org/abs/nucl-ex/0204008}{{\ttfamily nucl-ex/0204008}}].

\bibitem{Super-Kamiokande:1998kpq}
{\scshape Super-Kamiokande} collaboration, \emph{{Evidence for oscillation of
  atmospheric neutrinos}},
  \href{https://doi.org/10.1103/PhysRevLett.81.1562}{\emph{Phys. Rev. Lett.}
  {\bfseries 81} (1998) 1562}
  [\href{https://arxiv.org/abs/hep-ex/9807003}{{\ttfamily hep-ex/9807003}}].

\bibitem{Calibbi:2017uvl}
L.~Calibbi and G.~Signorelli, \emph{{Charged lepton flavour violation: an
  experimental and theoretical introduction}},
  \href{https://doi.org/10.1393/ncr/i2018-10144-0}{\emph{Riv. Nuovo Cim.}
  {\bfseries 41} (2018) 71}
  [\href{https://arxiv.org/abs/1709.00294}{{\ttfamily 1709.00294}}].

\bibitem{ParticleDataGroup:2022pth}
{\scshape Particle Data Group} collaboration, \emph{{Review of Particle
  Physics}},
  \href{https://doi.org/10.1093/ptep/ptac097}{\emph{PTEP} {\bfseries 2022}
  (2022) 083C01}.

\bibitem{Konetschny:1977bn}
W.~Konetschny and W.~Kummer, \emph{{Nonconservation of total lepton number with
  scalar bosons}},
  \href{https://doi.org/10.1016/0370-2693(77)90407-5}{\emph{Phys. Lett. B}
  {\bfseries 70} (1977) 433}.

\bibitem{Foot:1988aq}
R.~Foot, H.~Lew, X.G.~He and G.C.~Joshi, \emph{{See-saw neutrino masses induced
  by a triplet of leptons}},
  \href{https://doi.org/10.1007/BF01415558}{\emph{Z. Phys. C} {\bfseries 44}
  (1989) 441}.

\bibitem{Sun:2013kga}
K.-S.~Sun, T.-F.~Feng, G.-H.~Luo, X.-Y.~Yang and J.-B.~Chen, \emph{{Lepton
  flavor violation in inverse seesaw model}},
  \href{https://doi.org/10.1142/S0217732313501514}{\emph{Mod. Phys. Lett. A}
  {\bfseries 28} (2013) 1350151}
  [\href{https://arxiv.org/abs/1312.2073}{{\ttfamily 1312.2073}}].

\bibitem{Mohapatra:1980yp}
R.N.~Mohapatra and G.~Senjanovi\'c, \emph{{Neutrino masses and mixings in gauge
  models with spontaneous parity violation}},
  \href{https://doi.org/10.1103/PhysRevD.23.165}{\emph{Phys. Rev. D} {\bfseries
  23} (1981) 165}.

\bibitem{Hisano:1995nq}
J.~Hisano, T.~Moroi, K.~Tobe, M.~Yamaguchi and T.~Yanagida,
  \emph{{Lepton-flavor violation in the supersymmetric standard model with
  seesaw-induced neutrino masses}},
  \href{https://doi.org/10.1016/0370-2693(95)00954-J}{\emph{Phys. Lett. B}
  {\bfseries 357} (1995) 579}
  [\href{https://arxiv.org/abs/hep-ph/9501407}{{\ttfamily hep-ph/9501407}}].

\bibitem{Herrero-Garcia:2014usa}
J.~Herrero-Garcia, M.~Nebot, N.~Rius and A.~Santamaria, \emph{{Testing the
  Zee-Babu model via neutrino data, lepton flavour violation and direct
  searches at the LHC}},
  \href{https://doi.org/10.1016/j.nuclphysbps.2015.09.271}{\emph{Nucl. Part.
  Phys. Proc.} {\bfseries 273-275} (2016) 1678}
  [\href{https://arxiv.org/abs/1410.2299}{{\ttfamily 1410.2299}}].

\bibitem{Babu:1988ki}
K.S.~Babu, \emph{{Model of ``calculable'' Majorana neutrino masses}},
  \href{https://doi.org/10.1016/0370-2693(88)91584-5}{\emph{Phys. Lett. B}
  {\bfseries 203} (1988) 132}.

\bibitem{Zee:1980ai}
A.~Zee, \emph{{A theory of lepton number violation and neutrino Majorana
  masses}},
  \href{https://doi.org/10.1016/0370-2693(80)90349-4}{\emph{Phys. Lett. B}
  {\bfseries 93} (1980) 389}, [erratum:
  \href{https://doi.org/10.1016/0370-2693(80)90193-8}{\emph{Phys. Lett. B}
  {\bfseries 95} (1980) 461}].

\bibitem{Ma:2006km}
E.~Ma, \emph{{Verifiable radiative seesaw mechanism of neutrino mass and dark
  matter}},
  \href{https://doi.org/10.1103/PhysRevD.73.077301}{\emph{Phys. Rev.
  D} {\bfseries 73} (2006) 077301}
  [\href{https://arxiv.org/abs/hep-ph/0601225}{{\ttfamily hep-ph/0601225}}].

\bibitem{Rocha-Moran:2016enp}
P.~Rocha-Mor\'an and A.~Vicente, \emph{{Lepton flavor violation in the
  singlet-triplet scotogenic model}},
  \href{https://doi.org/10.1007/JHEP07(2016)078}{\emph{JHEP} {\bfseries 07}
  (2016) 078} [\href{https://arxiv.org/abs/1605.01915}{{\ttfamily
  1605.01915}}].

\bibitem{Hundi:2022iva}
R.S.~Hundi, \emph{{Lepton flavor violating $Z$ and Higgs decays in the
  scotogenic model}},
  \href{https://doi.org/10.1140/epjc/s10052-022-10453-3}{\emph{Eur. Phys. J. C}
  {\bfseries 82} (2022) 505}
  [\href{https://arxiv.org/abs/2201.03779}{{\ttfamily 2201.03779}}].

\bibitem{MEG:2016leq}
{\scshape MEG} collaboration, \emph{{Search for the lepton flavour violating
  decay $\mu^+ \rightarrow e^+ \gamma$ with the full dataset of the MEG
  experiment}},
  \href{https://doi.org/10.1140/epjc/s10052-016-4271-x}{\emph{Eur. Phys. J. C}
  {\bfseries 76} (2016) 434}
  [\href{https://arxiv.org/abs/1605.05081}{{\ttfamily 1605.05081}}].

\bibitem{SINDRUM:1987nra}
{\scshape SINDRUM} collaboration, \emph{{Search for the decay $\mu^+
  \rightarrow e^+ e^+ e^-$}},
  \href{https://doi.org/10.1016/0550-3213(88)90462-2}{\emph{Nucl. Phys. B}
  {\bfseries 299} (1988) 1}.

\bibitem{BaBar:2009hkt}
{\scshape BABAR} collaboration, \emph{{Searches for lepton flavor violation in
  the decays $\tau^\pm \rightarrow e^\pm \gamma$ and $\tau^\pm \rightarrow
  \mu^\pm \gamma$}},
  \href{https://doi.org/10.1103/PhysRevLett.104.021802}{\emph{Phys. Rev. Lett.}
  {\bfseries 104} (2010) 021802}
  [\href{https://arxiv.org/abs/0908.2381}{{\ttfamily 0908.2381}}].

\bibitem{Hayasaka:2010np}
K.~Hayasaka et~al., \emph{{Search for lepton-flavor-violating $\tau$ decays
  into three leptons with 719 million produced $\tau^+ \tau^-$ pairs}},
  \href{https://doi.org/10.1016/j.physletb.2010.03.037}{\emph{Phys. Lett. B}
  {\bfseries 687} (2010) 139}
  [\href{https://arxiv.org/abs/1001.3221}{{\ttfamily 1001.3221}}].

\bibitem{CMS:2016cvq}
{\scshape CMS} collaboration, \emph{{Search for lepton flavour violating decays
  of the Higgs boson to $e \tau$ and $e \mu$ in proton-proton collisions at
  $\sqrt s=$ 8 TeV}},
  \href{https://doi.org/10.1016/j.physletb.2016.09.062}{\emph{Phys. Lett. B}
  {\bfseries 763} (2016) 472}
  [\href{https://arxiv.org/abs/1607.03561}{{\ttfamily 1607.03561}}].

\bibitem{CMS:2017con}
{\scshape CMS} collaboration, \emph{{Search for lepton flavour violating decays
  of the Higgs boson to $\mu\tau$ and $e\tau$ in proton-proton collisions at
  $\sqrt{s}=$ 13 TeV}},
  \href{https://doi.org/10.1007/JHEP06(2018)001}{\emph{JHEP} {\bfseries 06}
  (2018) 001}
  [\href{https://arxiv.org/abs/1712.07173}{{\ttfamily 1712.07173}}].

\bibitem{OPAL:1995grn}
{\scshape OPAL} collaboration, \emph{{A search for lepton flavour violating
  $Z^0$ decays}},
  \href{https://doi.org/10.1007/BF01553981}{\emph{Z. Phys. C} {\bfseries 67}
  (1995) 555}.

\bibitem{DELPHI:1996iox}
{\scshape DELPHI} collaboration, \emph{{Search for lepton flavour number
  violating $Z^0$-decays}},
  \href{https://doi.org/10.1007/s002880050313}{\emph{Z. Phys. C} {\bfseries 73}
  (1997) 243}.

\bibitem{ATLAS:2022uhq}
{\scshape ATLAS} collaboration, \emph{{Search for the
  charged-lepton-flavor-violating decay $Z\rightarrow e\mu$ in $pp$ collisions
  at $\sqrt{s}=13$ TeV with the ATLAS detector}},
  \href{https://doi.org/10.1103/PhysRevD.108.032015}{\emph{Phys. Rev. D}
  {\bfseries 108} (2023) 032015}
  [\href{https://arxiv.org/abs/2204.10783}{{\ttfamily 2204.10783}}].

\bibitem{ZEUS:1996bbn}
{\scshape ZEUS} collaboration, \emph{{Search for lepton flavor violation in $e
  p$ collisions at 300 GeV center of mass energy}},
  \href{https://doi.org/10.1007/s002880050352}{\emph{Z. Phys. C} {\bfseries 73}
  (1997) 613} [\href{https://arxiv.org/abs/hep-ex/9704018}{{\ttfamily
  hep-ex/9704018}}].

\bibitem{H1:1999dil}
{\scshape H1} collaboration, \emph{{A search for leptoquark bosons and lepton
  flavor violation in $e^+ p$ collisions at HERA}},
  \href{https://doi.org/10.1007/s100529900195}{\emph{Eur. Phys. J. C}
  {\bfseries 11} (1999) 447}, [erratum:
  \href{https://doi.org/10.1007/s100520000361}{\emph{Eur. Phys. J. C}
  {\bfseries 14} (2000) 553}]
  [\href{https://arxiv.org/abs/hep-ex/9907002}{{\ttfamily hep-ex/9907002}}].

\bibitem{ZEUS:2002dnh}
{\scshape ZEUS} collaboration, \emph{{Search for lepton-flavor violation in
  $e^+ p$ collisions at DESY HERA}},
  \href{https://doi.org/10.1103/PhysRevD.65.092004}{\emph{Phys. Rev. D}
  {\bfseries 65} (2002) 092004}
  [\href{https://arxiv.org/abs/hep-ex/0201003}{{\ttfamily hep-ex/0201003}}].

\bibitem{ZEUS:2005nsy}
{\scshape ZEUS} collaboration, \emph{{Search for lepton-flavor violation at
  HERA}},
  \href{https://doi.org/10.1140/epjc/s2005-02399-1}{\emph{Eur. Phys. J. C}
  {\bfseries 44} (2005) 463}
  [\href{https://arxiv.org/abs/hep-ex/0501070}{{\ttfamily hep-ex/0501070}}].

\bibitem{H1:2007dum}
{\scshape H1} collaboration, \emph{{Search for lepton flavour violation in ep
  collisions at HERA}},
  \href{https://doi.org/10.1140/epjc/s10052-007-0440-2}{\emph{Eur. Phys. J. C}
  {\bfseries 52} (2007) 833}
  [\href{https://arxiv.org/abs/hep-ex/0703004}{{\ttfamily hep-ex/0703004}}].

\bibitem{H1:2011rlk}
{\scshape H1} collaboration, \emph{{Search for lepton flavour violation at
  HERA}},
  \href{https://doi.org/10.1016/j.physletb.2011.05.023}{\emph{Phys. Lett. B}
  {\bfseries 701} (2011) 20}
  [\href{https://arxiv.org/abs/1103.4938}{{\ttfamily 1103.4938}}].

\bibitem{FCC:2018byv}
{\scshape FCC} collaboration, \emph{{FCC physics opportunities}: {Future
  Circular Collider conceptual design report volume 1}},
  \href{https://doi.org/10.1140/epjc/s10052-019-6904-3}{\emph{Eur. Phys. J. C}
  {\bfseries 79} (2019) 474}.

\bibitem{LHeCStudyGroup:2012zhm}
{\scshape LHeC Study Group} collaboration, \emph{{A Large Hadron Electron
  Collider at CERN report on the physics and design concepts for machine and
  detector}}, \href{https://doi.org/10.1088/0954-3899/39/7/075001}{\emph{J.
  Phys. G} {\bfseries 39} (2012) 075001}
  [\href{https://arxiv.org/abs/1206.2913}{{\ttfamily 1206.2913}}].

\bibitem{Bruening:2013bga}
O.~Br\"uning and M.~Klein, \emph{{The Large Hadron Electron Collider}},
  \href{https://doi.org/10.1142/S0217732313300115}{\emph{Mod. Phys. Lett. A}
  {\bfseries 28} (2013) 1330011}
  [\href{https://arxiv.org/abs/1305.2090}{{\ttfamily 1305.2090}}].

\bibitem{LHeC:2020van}
{\scshape LHeC, FCC-eh Study Group} collaboration, \emph{{The Large
  Hadron-Electron Collider at the HL-LHC}},
  \href{https://doi.org/10.1088/1361-6471/abf3ba}{\emph{J. Phys. G} {\bfseries
  48} (2021) 110501} [\href{https://arxiv.org/abs/2007.14491}{{\ttfamily
  2007.14491}}].

\bibitem{Etesami:2021hex}
S.M.~Etesami, R.~Jafari, M.~Mohammadi Najafabadi and S.~Tizchang,
  \emph{{Searching for lepton flavor violating interactions at future
  electron-positron colliders}},
  \href{https://doi.org/10.1103/PhysRevD.104.015034}{\emph{Phys. Rev. D}
  {\bfseries 104} (2021) 015034}
  [\href{https://arxiv.org/abs/2107.00545}{{\ttfamily 2107.00545}}].

\bibitem{Calibbi:2021pyh}
L.~Calibbi, X.~Marcano and J.~Roy, \emph{{$Z$ lepton flavour violation as a
  probe for new physics at future $e^+ e^-$ colliders}},
  \href{https://doi.org/10.1140/epjc/s10052-021-09777-3}{\emph{Eur. Phys. J. C}
  {\bfseries 81} (2021) 1054}
  [\href{https://arxiv.org/abs/2107.10273}{{\ttfamily 2107.10273}}].

\bibitem{Altmannshofer:2023tsa}
W.~Altmannshofer, P.~Munbodh and T.~Oh, \emph{{Probing lepton flavor violation
  at circular electron-positron colliders}},
  \href{https://doi.org/10.1007/JHEP08(2023)026}{\emph{JHEP} {\bfseries 08}
  (2023) 026}
  [\href{https://arxiv.org/abs/2305.03869}{{\ttfamily 2305.03869}}].

\bibitem{Dev:2017ftk}
P.S.B.~Dev, R.N.~Mohapatra and Y.~Zhang, \emph{{Lepton flavor violation induced
  by a neutral scalar at future lepton colliders}},
  \href{https://doi.org/10.1103/PhysRevLett.120.221804}{\emph{Phys. Rev. Lett.}
  {\bfseries 120} (2018) 221804}
  [\href{https://arxiv.org/abs/1711.08430}{{\ttfamily 1711.08430}}].

\bibitem{Alan:2001cf}
A.T.~Alan and A.~Senol, \emph{{Lepton flavor changing neutral current processes
  at lepton hadron colliders}},
  \href{https://www.actaphys.uj.edu.pl/R/33/5/1343}{\emph{Acta Phys. Polon. B}
  {\bfseries 33} (2002) 1343}
  [\href{https://arxiv.org/abs/hep-ph/0110294}{{\ttfamily hep-ph/0110294}}].

\bibitem{Gonderinger:2010yn}
M.~Gonderinger and M.J.~Ramsey-Musolf, \emph{{Electron-to-tau lepton flavor
  violation at the Electron-Ion Collider}},
  \href{https://doi.org/10.1007/JHEP11(2010)045}{\emph{JHEP} {\bfseries 11}
  (2010) 045}, [erratum:
  \href{https://doi.org/10.1007/JHEP05(2012)047}{\emph{JHEP} {\bfseries 05}
  (2012) 047}]
  [\href{https://arxiv.org/abs/1006.5063}{{\ttfamily 1006.5063}}].

\bibitem{Accardi:2012qut}
A.~Accardi et~al., \emph{{Electron-Ion Collider: the next QCD frontier}:
  {understanding the glue that binds us all}},
  \href{https://doi.org/10.1140/epja/i2016-16268-9}{\emph{Eur. Phys. J. A}
  {\bfseries 52} (2016) 268}
  [\href{https://arxiv.org/abs/1212.1701}{{\ttfamily 1212.1701}}].

\bibitem{Cirigliano:2021img}
V.~Cirigliano, K.~Fuyuto, C.~Lee, E.~Mereghetti and B.~Yan, \emph{{Charged
  lepton flavor violation at the EIC}},
  \href{https://doi.org/10.1007/JHEP03(2021)256}{\emph{JHEP} {\bfseries 03}
  (2021) 256}
  [\href{https://arxiv.org/abs/2102.06176}{{\ttfamily 2102.06176}}].

\bibitem{AbdulKhalek:2021gbh}
R.~Abdul~Khalek et~al., \emph{{Science requirements and detector concepts for
  the Electron-Ion Collider}: {EIC yellow report}},
  \href{https://doi.org/10.1016/j.nuclphysa.2022.122447}{\emph{Nucl. Phys. A}
  {\bfseries 1026} (2022) 122447}
  [\href{https://arxiv.org/abs/2103.05419}{{\ttfamily 2103.05419}}].

\bibitem{AbdulKhalek:2022hcn}
R.~Abdul~Khalek et~al., \emph{{Snowmass 2021 White Paper: Electron Ion Collider
  for High Energy Physics}}, Contribution to
  \href{https://www.slac.stanford.edu/econf/C210711}{Snowmass 2021}
  [\href{https://arxiv.org/abs/2203.13199}{{\ttfamily 2203.13199}}].

\bibitem{Jueid:2023fgo}
A.~Jueid, J.~Kim, S.~Lee, J.~Song and D.~Wang, \emph{{Exploring lepton flavor
  violation phenomena of the $Z$ and Higgs bosons at electron-proton
  colliders}},
  \href{https://doi.org/10.1103/PhysRevD.108.055024}{\emph{Phys. Rev. D}
  {\bfseries 108} (2023) 055024}
  [\href{https://arxiv.org/abs/2305.05386}{{\ttfamily 2305.05386}}].

\bibitem{Illana:2000ic}
J.I.~Illana and T.~Riemann, \emph{{Charged lepton flavor violation from massive
  neutrinos in $Z$ decays}},
  \href{https://doi.org/10.1103/PhysRevD.63.053004}{\emph{Phys. Rev. D}
  {\bfseries 63} (2001) 053004}
  [\href{https://arxiv.org/abs/hep-ph/0010193}{{\ttfamily hep-ph/0010193}}].

\bibitem{DeRomeri:2016gum}
V.~De~Romeri, M.J.~Herrero, X.~Marcano and F.~Scarcella, \emph{{Lepton flavor
  violating $Z$ decays: a promising window to low scale seesaw neutrinos}},
  \href{https://doi.org/10.1103/PhysRevD.95.075028}{\emph{Phys. Rev. D}
  {\bfseries 95} (2017) 075028}
  [\href{https://arxiv.org/abs/1607.05257}{{\ttfamily 1607.05257}}].

\bibitem{ATLAS:2021bdj}
{\scshape ATLAS} collaboration, \emph{{Search for lepton-flavor violation in
  $Z$-boson decays with $\tau$-leptons with the ATLAS detector}},
  \href{https://doi.org/10.1103/PhysRevLett.127.271801}{\emph{Phys. Rev. Lett.}
  {\bfseries 127} (2021) 271801}
  [\href{https://arxiv.org/abs/2105.12491}{{\ttfamily 2105.12491}}].

\bibitem{Bauer:1977iq}
T.H.~Bauer, R.D.~Spital, D.R.~Yennie and F.M.~Pipkin, \emph{{The hadronic
  properties of the photon in high-energy interactions}},
  \href{https://doi.org/10.1103/RevModPhys.50.261}{\emph{Rev. Mod. Phys.}
  {\bfseries 50} (1978) 261}, [erratum: 
  \href{https://doi.org/10.1103/RevModPhys.51.407}{\emph{Rev. Mod. Phys.}
  {\bfseries 51} (1979) 407}].

\bibitem{Chwastowski:2003aw}
J.~Chwastowski and J.~Figiel, \emph{{Photoproduction at HERA}}, {\emph{Phys.
  Part. Nucl.} {\bfseries 35} (2004) 619}
  [\href{https://arxiv.org/abs/hep-ex/0311044}{{\ttfamily hep-ex/0311044}}].

\bibitem{Alloul:2013bka}
A.~Alloul, N.D.~Christensen, C.~Degrande, C.~Duhr and B.~Fuks, \emph{{FeynRules
  2.0 - a complete toolbox for tree-level phenomenology}},
  \href{https://doi.org/10.1016/j.cpc.2014.04.012}{\emph{Comput. Phys. Commun.}
  {\bfseries 185} (2014) 2250}
  [\href{https://arxiv.org/abs/1310.1921}{{\ttfamily 1310.1921}}].

\bibitem{Degrande:2011ua}
C.~Degrande, C.~Duhr, B.~Fuks, D.~Grellscheid, O.~Mattelaer and T.~Reiter,
  \emph{{UFO - the Universal FeynRules Output}},
  \href{https://doi.org/10.1016/j.cpc.2012.01.022}{\emph{Comput. Phys. Commun.}
  {\bfseries 183} (2012) 1201}
  [\href{https://arxiv.org/abs/1108.2040}{{\ttfamily 1108.2040}}].

\bibitem{Alwall:2011uj}
J.~Alwall, M.~Herquet, F.~Maltoni, O.~Mattelaer and T.~Stelzer, \emph{{MadGraph
  5: going beyond}},
  \href{https://doi.org/10.1007/JHEP06(2011)128}{\emph{JHEP} {\bfseries 06}
  (2011) 128} [\href{https://arxiv.org/abs/1106.0522}{{\ttfamily 1106.0522}}].

\bibitem{Sjostrand:2014zea}
T.~Sj\"ostrand, S.~Ask, J.R.~Christiansen, R.~Corke, N.~Desai, P.~Ilten et~al.,
  \emph{{An introduction to PYTHIA 8.2}},
  \href{https://doi.org/10.1016/j.cpc.2015.01.024}{\emph{Comput. Phys. Commun.}
  {\bfseries 191} (2015) 159}
  [\href{https://arxiv.org/abs/1410.3012}{{\ttfamily 1410.3012}}].

\bibitem{Buckley:2019xhk}
A.~Buckley, P.~Ilten, D.~Konstantinov, L.~L\"onnblad, J.~Monk, W.~Pokorski
  et~al., \emph{{The HepMC3 event record library for Monte Carlo event
  generators}},
  \href{https://doi.org/10.1016/j.cpc.2020.107310}{\emph{Comput. Phys. Commun.}
  {\bfseries 260} (2021) 107310}
  [\href{https://arxiv.org/abs/1912.08005}{{\ttfamily 1912.08005}}].

\bibitem{deFavereau:2013fsa}
{\scshape DELPHES 3} collaboration, \emph{{DELPHES 3: a modular framework for
  fast simulation of a generic collider experiment}},
  \href{https://doi.org/10.1007/JHEP02(2014)057}{\emph{JHEP} {\bfseries 02}
  (2014) 057}
  [\href{https://arxiv.org/abs/1307.6346}{{\ttfamily 1307.6346}}].

\bibitem{Cacciari:2011ma}
M.~Cacciari, G.P.~Salam and G.~Soyez, \emph{{FastJet user manual}},
  \href{https://doi.org/10.1140/epjc/s10052-012-1896-2}{\emph{Eur. Phys. J. C}
  {\bfseries 72} (2012) 1896}
  [\href{https://arxiv.org/abs/1111.6097}{{\ttfamily 1111.6097}}].

\bibitem{Roe:2004na}
B.P.~Roe, H.-J.~Yang, J.~Zhu, Y.~Liu, I.~Stancu and G.~McGregor, \emph{{Boosted
  decision trees as an alternative to artificial neural networks}},
  \href{https://doi.org/10.1016/j.nima.2004.12.018}{\emph{Nucl. Instrum. Meth.
  A} {\bfseries 543} (2005) 577}
  [\href{https://arxiv.org/abs/physics/0408124}{{\ttfamily physics/0408124}}].

\bibitem{Chen:2016btl}
T.~Chen and C.~Guestrin, \emph{{XGBoost: a scalable tree boosting system}}, in
  Proceedings of the 22nd ACM SIGKDD International Conference on Knowledge
  Discovery and Data Mining, publisher: Association for Computing Machinery,
  2016, p.~785,
  doi:\href{https://doi.org/10.1145/2939672.2939785}{10.1145/2939672.2939785}
  [\href{https://arxiv.org/abs/1603.02754}{{\ttfamily 1603.02754}}].

\bibitem{Dey:2020tfq}
A.~Dey, J.~Lahiri and B.~Mukhopadhyaya, \emph{{LHC signals of triplet scalars
  as dark matter portal: cut-based approach and improvement with gradient
  boosting and neural networks}},
  \href{https://doi.org/10.1007/JHEP06(2020)126}{\emph{JHEP} {\bfseries 06}
  (2020) 126}
  [\href{https://arxiv.org/abs/2001.09349}{{\ttfamily 2001.09349}}].

\bibitem{Dam:2018rfz}
M.~Dam, \emph{Tau-lepton physics at the FCC-ee circular $e^+ e^-$ collider},
  \href{https://doi.org/10.21468/SciPostPhysProc.1.041}{\emph{SciPost Phys. Proc.}
  {\bfseries 1} (2019) 041}
  [\href{https://arxiv.org/abs/1811.09408}{{\ttfamily 1811.09408}}].

\end{thebibliography}

\providecommand{\href}[2]{#2}\begingroup\raggedright\endgroup

\end{document}